\documentclass[useAMS]{mn2e}

\usepackage{psfig,epsfig,supertabular,wrapfig,placeins}
\usepackage{graphicx,amsmath,amssymb,subfigure,multirow,units}
\usepackage{marvosym}
\usepackage{graphics}
\usepackage{amssymb}
\usepackage{amsmath}
\usepackage{euscript}
\usepackage{setspace}
\usepackage{fancyhdr}
\usepackage{afterpage,rotating,url}
\usepackage{times}

\newcommand{\asec}{\ensuremath{^{\prime\prime}}}

\newcommand{\gtsim}{\mbox{{\raisebox{-0.4ex}{$\stackrel{>}{{\scriptstyle\sim}}$}
}}}
\newcommand{\ltsim}{\mbox{{\raisebox{-0.4ex}{$\stackrel{<}{{\scriptstyle\sim}}$}
}}}

\title[{\em Herschel}--ATLAS: Counterparts]{{\em Herschel}--ATLAS:
  counterparts from the UV--NIR in the science demonstration phase
  catalogue\thanks{{\em Herschel} is an ESA space observatory with
    science instruments provided by European-led Principal
    Investigator consortia and with important participation from
    NASA}}

\author[D.J.B.~Smith et al.]
{D.J.B. Smith$^{1,2}$\thanks{E-mail: daniel.j.b.smith@gmail.com}
~L. Dunne$^{1}$, 
~S.J. Maddox$^{1}$,
~S. Eales$^3$, % Cardiff = 2
~D.G. Bonfield$^{2}$, %Herts = 2
~M.J.~Jarvis$^{2}$, %Herts 
\newauthor
W.~Sutherland$^4$, % QMUW = 4
~S.~Fleuren$^4$, % QMUW
~E.E.~Rigby$^1$,% Nottingham
~M.A.~Thompson$^{2}$, % Herts
~I.K.~Baldry$^5$, %LJMU = 5
\newauthor
S.~Bamford$^1$, % Notts
~S.~Buttiglione$^6$,  % INAF = 6
~A.~Cava$^{28}$, % IAC= 8
~D.L.~Clements$^8$, % Imperial - 9  
~A.~Cooray$^{9}$, % UC IRVINE - 10     
~S.~Croom$^{10}$, % Sydney = 11
\newauthor 
A.~Dariush$^3$,% Cardiff
~G.~de Zotti$^{6,11}$, %- INAF (6), and SISSA (12) 
~S.P.~Driver$^{12,29}$,   % SUPA St Andrews = 13
~J.S.~Dunlop$^{13}$, % SUPA ROE, IoA - 14
~J.~Fritz$^{14}$, % Gent = 15
~D.T.~Hill$^{12}$,   % SUPA St Andrews 
\newauthor 
A.~Hopkins$^{15}$, % AAO Australia = 16
~R.~Hopwood$^{16}$, % OPEN UNI - 17 
~E.~Ibar$^{17}$, % UK ATC, Edinburgh = 18
~R.J.~Ivison$^{17,13}$, % UK ATC, Edinburgh, and IoA
~D.H.~Jones$^{15}$, % AAO Australia
\newauthor
L.~Kelvin$^{12}$, % SUPA St Andrews
~L.~Leeuw$^{18}$,  % -- SETI - 19     
~J.~Liske$^{19}$, % ESO, Garching = 20
~J.~Loveday$^{20}$, % Sussex = 21
~B.F.~Madore$^{21}$, % Carnegie Observatories = 22
~P.~Norberg$^{13}$, % SUPA, ROE 
\newauthor
P.~Panuzzo$^{22}$,  % CEA Saclay = 23
~E.~Pascale$^3$,% Cardiff
~M.~Pohlen$^3$,% Cardiff
~C.C.~Popescu$^{23}$,  % UCLAN = 24
~M.~Prescott$^{5}$, % LJMU
~A.~Robotham$^{12}$, % SUPA St Andrews
\newauthor
G.~Rodighiero$^{6}$, %INAF Italy
~D.~Scott$^{24}$, %UBC - 25    
~M.~Seibert$^{21}$, %  Observatories of Carnegie
~R.~Sharp$^{15}$, % AAO Australia
~P.~Temi$^{25}$, %NASA AMES = 26
~R.J.~Tuffs$^{26}$, % MPIK = 27
\newauthor
P.~van der Werf$^{27,13}$, % leiden = 28     
~E.~van~Kampen$^{19}$ % ESO
  \\$^{1}$Centre for Astronomy and Particle Theory, The School of
  Physics \&\ Astronomy, Nottingham University, University Park
  Campus, Nottingham,\\NG7 1HR, UK \\ 
  $^{2}$Centre for Astrophysics Research, Science \&\ Technology Research
  Institute, University of Hertfordshire, Hatfield, Herts, AL10 9AB,
  UK \\ 
  $^3$School of Physics \&\ Astronomy, Cardiff University, Queen
  Buildings, The Parade, Cardiff, CF24 3AA, UK \\
  $^4$School of Mathematical Sciences, Queen Mary, University of
  London, Mile End Road, London, E1 4NS, UK\\
  $^5$Astrophysics Research Institute, Liverpool John Moores University, Twelve Quays House, Egerton Wharf, Birkenhead, CH41 1LD, UK\\
  $^6$INAF -- Osservatorio Astronomico di Padova, Vicolo Osservatorio 5, I-35122, Padova, Italy\\
  $^7$Instituto de Astrof\'isica de Canarias (IAC) and Departamento de
  Astrof\'isica de La Laguna (ULL), La Laguna, Tenerife, Spain\\
  $^8$Astrophysics Group, Imperial College London, Blackett Laboratory, Prince Consort Road, London SW7 2AZ, UK\\
  $^{9}$University of California, Ivine, Department of Physics \&\ Astronomy, 4186 Frederick Reines Hall, Irvine, CA 92697-4575, USA\\
  $^{10}$Sydney Institute for Astronomy, School of Physics, University of Sydney, NSW 2006, Australia\\
  $^{11}$SISSA, Via Bonomea 265, I-34136 Trieste, Italy\\
  $^{12}$SUPA, School of Physics and Astronomy, University of St. Andrews, North Haugh, St. Andrews, KY16 9SS, UK\\
  $^{13}$Institute for Astronomy, University of Edinburgh, Royal Observatory, Edinburgh, EH9 3HJ, UK\\
  $^{14}$Sterrenkundig Observatorium, Universiteit Gent, Krijgslaan 281 S9, B-9000 Gent, Belgium\\
  $^{15}$Anglo-Australian Observatory, PO Box 296, Epping, NSW 1710, Australia\\
  $^{16}$Department of Physics and Astronomy, The Open University, Walton Hall, Milton Keynes MK7 6AA, UK\\
  $^{17}$Uk Astronomy Technology Centre, Royal Observatory, Edinburgh, EH9 3HJ, UK\\
  $^{18}$SETI Institute, 515 N. Whisman Avenue, Mountain View, CA 94043, USA\\
  $^{19}$European Southern Observatory, Karl-Schwarzschild-Strasse 2, D-85748, Garching bei M\"unchen, Germany\\
  $^{20}$Astronomy Centre, University of Sussex, Falmer, Brighton, BN1 9QH, UK\\
  $^{21}$Observatories of the Carnegie Institution, 813 Santa Barbera St., Pasadena, CA 91101, USA\\
  $^{22}$CEA, Laboratoire AIM, Irfu\slash SAp, Orme des Merisiers, F-91191 Gif-sur-Yvette, France\\
  $^{23}$Jeremiah Horrocks Institute, University of Central Lancashire, Preston, PR1 2HE, UK\\
  $^{24}$Department of Physics and Astronomy, 6224 Agricultural Road, University of British Columbia, Vancouver, BC, V6T 1Z1, Canada\\
  $^{25}$Astrophysics Branch, NASA Ames Research Center, Mail Stop 2456,  Moffett Field, CA 94035, USA\\
  $^{26}$Max Planck Institut f\"ur Kernphysik (MPIK), Saupfercheckweg, 69117 Heidelberg, Germany\\
  $^{27}$Leiden Observatory, Leiden University, PO Box 9513, NL - 2300 RA Leiden, The Netherlands\\
  $^{28}$Departamento de Astrof\'{\i}sica, Facultad de CC. F\'{\i}sicas, Universidad Complutense de Madrid, E-28040 Madrid, Spain\\
  $^{29}$International Centre for Radio Astronomy Research, The University of Western Australia, 7 Fairway, Crawley, Perth, Western Australia, WA6009
 }
\begin{document}

\date{\today}

\pagerange{\pageref{firstpage}--\pageref{lastpage}} \pubyear{2010}

\maketitle

\label{firstpage}
\clearpage

\begin{abstract}
We present a technique to identify optical counterparts of
250\,$\mu$m-selected sources from the {\em Herschel}-ATLAS survey. Of
the 6621 250\,$\mu$m $> 32$\,mJy sources in our science demonstration
catalogue we find that $\sim 60$ percent have counterparts brighter
than $r=22.4$\,mag in the Sloan Digital Sky Survey. Applying a
likelihood ratio technique we are able to identify 2423 of the
counterparts with a reliability $R > 0.8$. This is approximately 37
percent of the full 250\,$\mu$m catalogue. We have estimated
photometric redshifts for each of these 2423 reliable counterparts,
while 1099 also have spectroscopic redshifts collated from several
different sources, including the GAMA survey.  We estimate the
completeness of identifying counterparts as a function of redshift,
and present evidence that 250\,$\mu$m-selected {\it Herschel}-ATLAS
galaxies have a bimodal redshift distribution.  Those with reliable
optical identifications have a redshift distribution peaking at $z
\approx 0.25 \pm 0.05$, while sub-mm colours suggest that a
significant fraction with no counterpart above the r-band limit have
$z > 1$. We also suggest a method for selecting populations of
strongly-lensed high redshift galaxies. Our identifications are
matched to UV--NIR photometry from the GAMA survey, and these data are
available as part of the {\it Herschel}-ATLAS public data release.
\end{abstract}


\section{Introduction}
One of the key problems to overcome when conducting multi-wavelength
surveys is determining which sources are associated with one another
in different wave-bands, and which are unrelated. When multiple
observations have been conducted at similar wavelengths and with
similar resolution and sensitivity, this problem can be reliably
addressed by using a simple nearest--neighbour match. However, in the
situation where the two distinct sets of observations to be matched
have considerably different resolution -- for example matching
far-infrared or sub-millimetre survey data to an optical catalogue
(e.g. Sutherland et al., 1991, Clements et al., 1996, Serjeant et al.,
2003, Clements et al., 2004, Ivison et al. 2005, 2007, Wang
\&\ Rowan-Robinson, 2009, Biggs et al. 2010) -- the large positional
uncertainties in the longer-wavelength data can make it much more
difficult to find reliable associations between sub--millimetre
sources and their optical\slash near--infrared counterparts.

One method which can be used to identify the most likely counterpart
to a low-resolution source, is the Likelihood Ratio technique
(hereafter LR), first suggested by Richter (1975), and expanded by
Sutherland \&\ Saunders (1992) and Ciliegi et al. (2003). The crucial
advantage of the LR technique over other methods is that it not only
uses the positional information contained within the two catalogues,
but also includes brightness information (both of the individual
potential counterparts, and of the higher resolution catalogue as a
whole) to identify the most reliable counterpart to a low-resolution
source.

The {\em Herschel} Astrophysical Terahertz Large Area Survey ({\em
  Herschel}--ATLAS, Eales et al., 2010) is the largest open-time key
project that will be carried out with the {\em Herschel Space
  Observatory} (Pilbratt et al., 2010). The {\em Herschel}--ATLAS will
survey in excess of 550 deg$^2$ in five channels centred on 100, 160,
250, 350 and 500\,$\mu$m, using the PACS (Poglitsch et al., 2010) and
SPIRE instruments (Griffin et al., 2010). This makes {\em
  Herschel}--ATLAS currently the largest area extragalactic {\em
  Herschel} survey.  The {\em Herschel}--ATLAS observations consist of
two scans in parallel mode reaching 5$\sigma$ point source
sensitivities of 132, 126, 32, 36 and 45 mJy in the 100\,$\mu$m,
160\,$\mu$m, 250\,$\mu$m, 350\,$\mu$m and 500\,$\mu$m channels
respectively, with beam sizes of approximately 9, 13, 18, 25 and 35
arcsec in the same five bands. The SPIRE and PACS map-making
procedures are described in the papers by Pascale et al. (2010) and
Ibar et al. (2010), while the catalogues are described in Rigby et
al. (2010). One of the primary aims of the {\em Herschel}--ATLAS was
to obtain the first unbiased survey of the local Universe at sub-mm
wavelengths, and as a result the survey was designed to overlap with
existing large optical and infrared surveys.

In this paper, we present a discussion of our implementation of the LR
technique to identify the most reliable counterparts to
250\,$\mu$m--selected sources in the {\em Herschel}--ATLAS science
demonstration phase (SDP) data field (Eales et al., 2010). This field
was chosen in order to take advantage of multi--wavelength data from
the Sloan Digital Sky Survey (SDSS -- York et al., 2000), and the UK
Infrared Deep Sky Survey Large Area Survey (UKIDSS-LAS -- Lawrence et
al., 2007). This field also overlaps with the 9 hour field of the
Galaxy And Mass Assembly survey (GAMA -- Driver et al., 2010). The
GAMA catalogue (Hill et al., 2011), comprises not only thousands of
redshifts (for galaxies selected as described in Baldry et al., 2010,
and observed with the maximum possible tiling efficiency -- Robotham
et al., 2010), but also $r$--band--defined aperture--matched
photometry in the $ugrizYJHK$ bands. In addition, the GAMA fields are
being systematically observed using the {\em Galaxy Evolution EXplorer
  (GALEX)} satellite (Martin et al., 2005) at Medium Imaging Survey
depth to provide aperture--matched FUV and NUV counterparts to the
catalogued GAMA sources (the {\em GALEX}--GAMA survey; Seibert et al.,
in prep). These counterparts will potentially be of great scientific
value once the most reliable optical counterpart can be established
for each {\em Herschel}--ATLAS source.

In section \ref{LRcalc} we present the specific LR method that we have
used to identify counterparts to 250\,$\mu$m--selected sources from
the {\em Herschel}--ATLAS SDP catalogue in an $r$--band catalogue of
model magnitudes derived from the SDSS DR7. In section
\ref{sec:redshifts} we present the redshift properties of our
catalogue, which covers $\sim 16$\,deg$^2$ over the GAMA 9 hour
field. Section \ref{results} contains some basic results based on our
reliable catalogue, and in section \ref{conclusions} we present some
concluding remarks about the likelihood ratio technique and the
resulting catalogue.

\section{Calculating the Likelihood Ratio}
\label{LRcalc}

The likelihood ratio, i.e. the ratio between the probability that the
source is the correct identification and the corresponding probability
for an unrelated background source, is calculated as in Sutherland \&
Saunders (1992):

\begin{equation}
L = \frac{q(m)f(r)}{n(m)},
\end{equation}

\noindent in which $n(m)$, and $q(m)$ correspond to the SDSS $r$--band
magnitude probability distributions of the full $r$--band catalogue
and of the true counterparts to the sub--millimetre sources,
respectively, while $f(r)$ represents the radial probability
distribution of offsets between the 250\,$\mu$m positions and the SDSS
$r$--band centroids. We will now describe how we calculate each
component of this relationship in turn.

\subsection{Calculating the radial dependence of the likelihood ratio, $f(r)$}
\label{sec:f_r}

Here, $f(r)$ is the radial probability distribution function of the
positional errors as a function of the separation from the SPIRE
250\,$\mu$m position in arcseconds ($r$), given by:

\begin{equation}
f(r) = \frac{1}{2\pi\sigma^2_{\rm pos}} \exp \left(
\frac{-r^2}{2\sigma_{\rm pos}^2} \right)
\label{eq:f_r}
\end{equation}

\noindent where $r$ is the separation between the 250\,$\mu$m and
$r$--band positions, and $\sigma_{\rm pos}$ is the standard positional
error (which is assumed to be isotropic).

\begin{figure}
\centering \includegraphics[width=0.99\columnwidth]{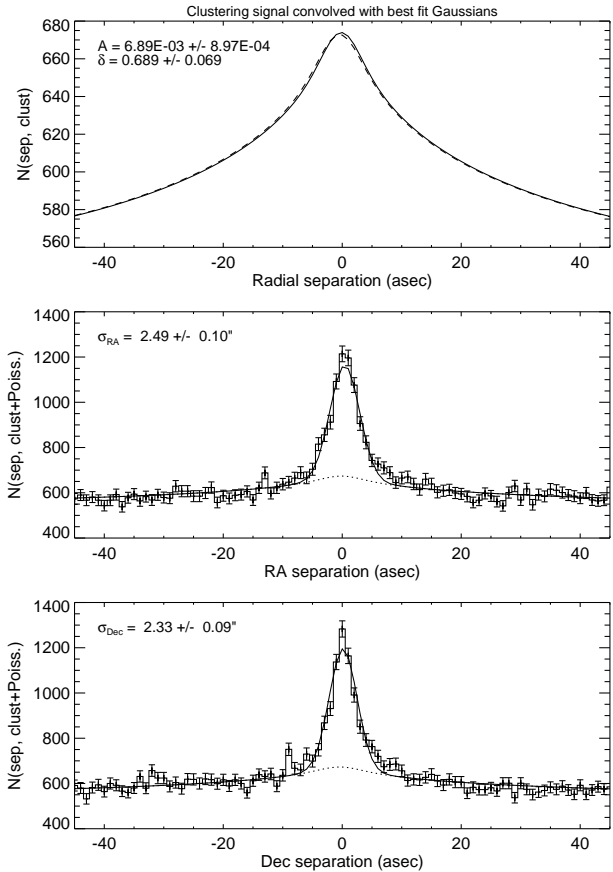}
\caption{In order to derive the 1$\sigma$ positional errors of the
  SPIRE 250\,$\mu$m--selected sources, we produced histograms of the
  total number of SDSS sources within a box 50 arcsec on a side around
  the SPIRE 250\,$\mu$m centres. After accounting for the clustering
  of SDSS sources (the top panel shows the signal expected for the
  clustering of SDSS sources in the RA and Dec directions convolved
  with Gaussian positional errors -- the results are shown as solid
  and dashed lines for RA and Dec, respectively, and these results
  appear as the dotted lines in the bottom two panels), we can add in
  an appropriate Gaussian distribution of centres to account for the
  actual positions of the SPIRE sources (equations \ref{xhist}
  \&\ \ref{yhist}). Performing a $\chi^2$ minimisation allows us to
  then empirically determine the 1$\sigma$ positional uncertainty for
  these sources, which are shown as $\sigma_{\rm RA}$ and $\sigma_{\rm
    Dec}$.}
\label{fig:clust}
\end{figure}

For {\em Herschel}--ATLAS SDP observations, it was necessary to
determine the SPIRE positional uncertainties. Since this information
was not available {\it a priori}, we empirically estimated
$\sigma_{\rm pos}$ using the SDSS DR7 $r$--band catalogue positions,
assuming that the SDSS positional errors were negligible in comparison
to the SPIRE errors. To determine $\sigma_{\rm pos}$, we derived
histograms of the separations between the positions in the MAD-X SPIRE
catalogue (Rigby et al., 2010) of the $5\sigma$ 250\,$\mu$m sources,
and all of those objects in the $r$--band SDSS DR7 catalogue within 50
arcsec, doing this for both the North--South and East--West directions
(Figure \ref{fig:clust}). These histograms can be well--described as
the sum of the Gaussian positional errors plus the clustering signal
for SDSS sources convolved with Gaussian errors, $G(\theta, \sigma)$,
with $\sigma = \sigma_{\rm pos}$:

\begin{equation}
n(x) = G^\prime (x,\sigma_{\rm pos}) + \left(\sum_yw(\theta) *
G(\theta,\sigma_{\rm pos})\right),
\label{xhist}
\end{equation}

\begin{equation}
n(y) = G^\prime(y,\sigma_{\rm pos}) + \left(\sum_xw(\theta) *
G(\theta,\sigma_{\rm pos})\right),
\label{yhist}
\end{equation} 

\noindent where $w(\theta) = A\theta^{-\delta}$, with $\theta$ being
measured in degrees for the purposes of comparison with the
literature. We determined the values of $A$ and $\delta$ empirically
based solely on galaxies in the SDSS catalogue over $>35$ deg$^2$
centred on the {\em Herschel}--ATLAS SDP field (limited to $r <
22.4$), with the
%best fit parameters $A = 2.01 \pm 0.26 \times 10^{-3}$ and $\delta =
%0.803 \pm 0.087$, 
best fit parameters $A = 6.89 \pm 0.90 \times 10^{-3}$ and $\delta =
0.689 \pm 0.069$, in reasonable agreement with the values of Connolly
et al., 2002. The effects of clustering (i.e. $w(\theta) *
G(\theta,\sigma_{\rm pos})$) are shown in the top panel of Figure
\ref{fig:clust}.

In order to determine the 1$\sigma$ positional error of the
250\,$\mu$m selected catalogue, we conducted a simple $\chi^2$ fit of
our model (equations \ref{xhist} \& \ref{yhist}) to the
histograms. The results are shown in Figure \ref{fig:clust} for the
summations in the East-West and North-South directions in the middle
and bottom panels respectively. The clustering signal is shown in the
bottom two panels by the dotted lines, with the histograms and their
Poisson error bars overlaid with the best fit model (solid lines). The
1$\sigma$ positional errors were found to be $2.49\pm 0.10$ arcsec and
$2.33 \pm 0.09$ arcsec in the two directions, consistent with one
another within the errors. The advantages of this method are
two--fold; firstly, it is not necessary to identify the counterparts
to the 250\,$\mu$m sources {\it a priori}, and secondly, the centroids
of the best fit Gaussians may be used to determine astrometric
corrections in the SPIRE maps (e.g. Pascale et al., 2010). The value
for $\sigma_{\rm pos}$ that we adopted was the weighted mean $2.40 \pm
0.09$ arcsec.

Theoretically, the positional uncertainty should depend on the
signal-to-noise ratio (SNR) of the detection and on the full-width at
half maximum (FWHM) of the SPIRE 250 \,$\mu$m beam (18.1 arcsec, Pascale
et al., 2010), following the results derived in Ivison et al. (2007;
$\sigma_{\rm th} = 0.6 \frac{\rm FWHM}{\rm SNR}$) and assuming the
case of uncorrelated noise. We use our empirical results in Figure
\ref{fig:clust} to calibrate the theoretical relation presented in
Ivison et al. (2007) to our data, and assume that our results are
symmetric in RA and Dec. This leads us to introduce a factor of 1.09
to give equation \ref{ivison2007}:

\begin{align}
\sigma_{\rm pos} &= 1.09 \times \sigma_{\rm th} \\ 
 &= 0.655 \frac{\rm FWHM}{\rm SNR}.
\label{ivison2007}
\end{align}

%We also note that if we limit ourselves to SPIRE sources of higher
%signal to noise (e.g. $> 10\sigma$ sources), the empirical estimates
%for $\sigma_{\rm pos}$ are consistent with our modified equation
%\ref{ivison2007}.

Although the SNR of some SPIRE 250\,$\mu$m sources is very high, it is
unphysical to allow $\sigma_{\rm pos}$ in equation \ref{ivison2007} to
approach zero for three main reasons:
\begin{itemize}
\item Whilst it is acceptable to neglect the SDSS DR7 positional
  errors for the purposes of determining $\sigma_{\rm pos}$ (section
  \ref{sec:f_r}), the astrometric precision for sources in the SDSS
  DR7 catalogue is non-zero ($< 0.1$ arcsec -- Abazajian et al.,
  2009).
\item Large sources, especially those without Gaussian surface
  brightness profiles (e.g. bright spiral galaxies), have considerably
  larger positional uncertainties associated with them.
\item Confusion provides a lower limit to the positional errors of the
  SPIRE catalogue, although the SNR in equation \ref{ivison2007} does
  include confusion noise as described in Rigby et al. (2010) and
  Pascale et al (2010).
\end{itemize}

\noindent Other effects that can influence the positional uncertainty
include imprecise knowledge of the beam morphology and the effects of
drifts and jitter in the {\em Herschel} pointing model.

To account for these effects, we do not allow the positional
uncertainty to fall below 1 arcsec, and we also include a term which
adds 5 percent of the SDSS $r$--band isophotal major axis in
quadrature to the value determined by equation \ref{ivison2007}, for
those sources with $r$--band model magnitudes $< 20.5$. Finally,
$f(r)$ must be renormalised so that

\begin{equation}
2\pi\int_0^\infty f(r) r dr = 1.
\end{equation}

\subsection{Calculating the magnitude dependence of the likelihood ratio}
\label{sec:lr_magdep}

Calculating the LR requires two further pieces of magnitude
information, $n(m)$ and $q(m)$. The quantity $n(m)$ is simply the
probability that a background source is observed with magnitude
$m$. To estimate this, we calculate the distribution of SDSS DR7
$r$--band model magnitudes for all of the primary photometry sources
in the catalogue, normalised to the total area of the catalogue (which
is approximately 36.0 deg$^2$ for the SDSS catalogue that we use for
this purpose). 

The non-triviality lies in the calculation of $q(m)$ -- the
probability that a true counterpart to a 250$\mu$m source has a
magnitude $m$. To estimate this we calculate the $r$--band magnitude
distribution of the counterparts to the 250\,$\mu$m sources using the
method of Ciliegi et al. (2005). This method involves counting all
objects in the optical catalogue within some fixed maximum search
radius ($r_{\rm max}$) of the SPIRE positions. To avoid influencing
the results of this analysis with erroneous deblends in the SDSS DR7
catalogue (which artificially alter the number counts), we eyeballed
the SDSS $r$--band images of each of the 5$\sigma$ 250\,$\mu$m
sources, removing 370 SDSS sources from the input catalogue. The
magnitude distribution of the remaining objects is referred to as
total$(m)$. Here we have adopted $r_{\rm max} = 10$\,arcsec, which
encloses $>$99.996\%\ of the real counterparts to the 250\,$\mu$m
sources based on our derived value for $\sigma_{\rm pos}$. The
distribution total$(m)$ is then background--subtracted to leave the
magnitude distribution of excess sources around the 250\,$\mu$m
centres, real$(m)$:

\begin{equation}
{\rm real}(m) = \left[{\rm total}(m) - \left(n(m) \times N_{\rm
    centres} \times \pi \times r_{\rm max}^2\right) \right],
\end{equation}

\noindent where $N_{\rm centres}$ is the number of 250\,$\mu$m sources
in the catalogue. This enables us to empirically estimate $q(m)$ from
the sources in our optical catalogue rather than modelling the
$r$--band magnitude distribution of 250\,$\mu$m--selected {\em
  Herschel}--ATLAS sources. The distribution $q(m)$ is given by
equation \ref{eq:q_m}:

\begin{equation}
q(m) = \frac{{\rm real}(m)}{\sum_m{\rm real}(m)}\times Q_0.
\label{eq:q_m}
\end{equation}

\noindent $Q_0$ is the fraction of true counterparts which are above
the SDSS limit, and is calculated thus:

\begin{equation}
Q_0 = \frac{N_{\rm matches} - \left(\sum_m n(m) \times \pi r_{\rm
    max}^2 \times N_{\rm centres}\right)}{N_{\rm centres}},
\label{eq:q0}
\end{equation}

\noindent here $N_{\rm matches}$ represents the number of possible IDs
within 10.0 arcsec of the SPIRE positions, and $N_{\rm centres}$ is
defined as above.  Since the value of $Q_0$ will be different for
galaxies and unresolved sources in our catalogue, we must calculate
$q(m)$, $n(m)$ and $Q_0$ separately for each population.

\begin{figure}
\centering \includegraphics[width=0.99\columnwidth]{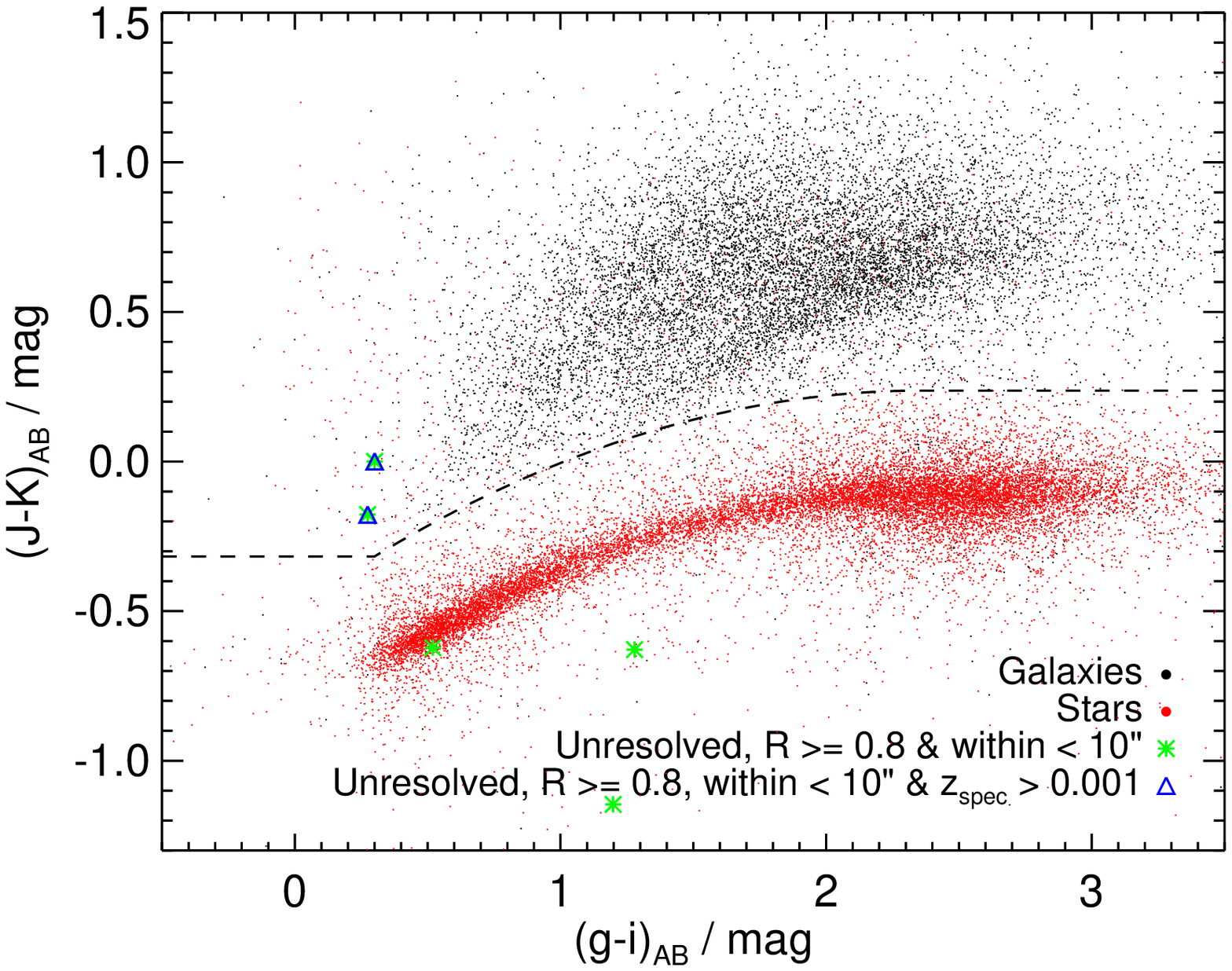}
\caption{Colour--colour diagram for sources in the GAMA catalogue. We
  use the relationship of Baldry et al. (2010) to distinguish between
  unresolved sources (stars and QSOs) and galaxies. Those sources with
  stellar\slash QSO colours in the SDSS\slash LAS catalogue data over the
  GAMA 9 hour field are displayed in red, while those with galaxy
  colours are displayed in black. Objects with colours consistent with
  QSOs are located toward the upper left corner of this plot. Of the
  five $R \ge 0.8$ sources in our catalogue classified as unresolved,
  we find that three satisfy the GAMA colour selection criteria for
  being stellar, and so are potentially evolved stars, dust-obscured
  QSOs or debris disk candidates possibly indicative of a
  proto-planetary system (e.g. Thompson et al., 2010). The dashed line
  describes the first order star--galaxy separation locus (for more
  details see Baldry et al., 2010). The star--galaxy separation locus
  has been modified slightly from the Baldry et al. value due to the
  fainter magnitudes considered in our survey. }
\label{fig:stargalsep}
\end{figure}

We separate resolved and unresolved sources using a slightly modified
version of the GAMA colour--colour relation from Baldry et al. (2010,
modified such that $\Delta_{sg,jk} > 0.40$ rather than 0.20 to avoid
adding an unphysical sharp edge to the stellar locus in Figure
\ref{fig:stargalsep}). Having separated the two populations, we
corrected the positions of the unresolved sources for known proper
motions in the USNO\slash SDSS DR7 catalogue (Munn et al., 2004),
precessing their co--ordinates to the epoch of the {\em
  Herschel}--ATLAS SDP observations. Only those unresolved sources
with proper motions detected at a SNR $\ge 3$ were updated.

For our SDSS DR7 $r$--band catalogue, $Q_0^{\rm gal} = 0.583$,
i.e. 58.3 percent of the galaxy counterparts are brighter than our
magnitude limit. For the unresolved sources the value is $Q_0^{\rm
  unres} = 0.010$, indicating that only 1 percent of the unresolved
sources in the catalogue are detected at $\ge 5\sigma$ in our
250$\mu$m data (although see section \ref{sec:randcat}). Thus we
determine that overall $Q_0 = Q_0^{\rm gal} + Q_0^{\rm unres} =
0.593$.

\begin{figure*}
\centering
\includegraphics[width=0.95\textwidth]{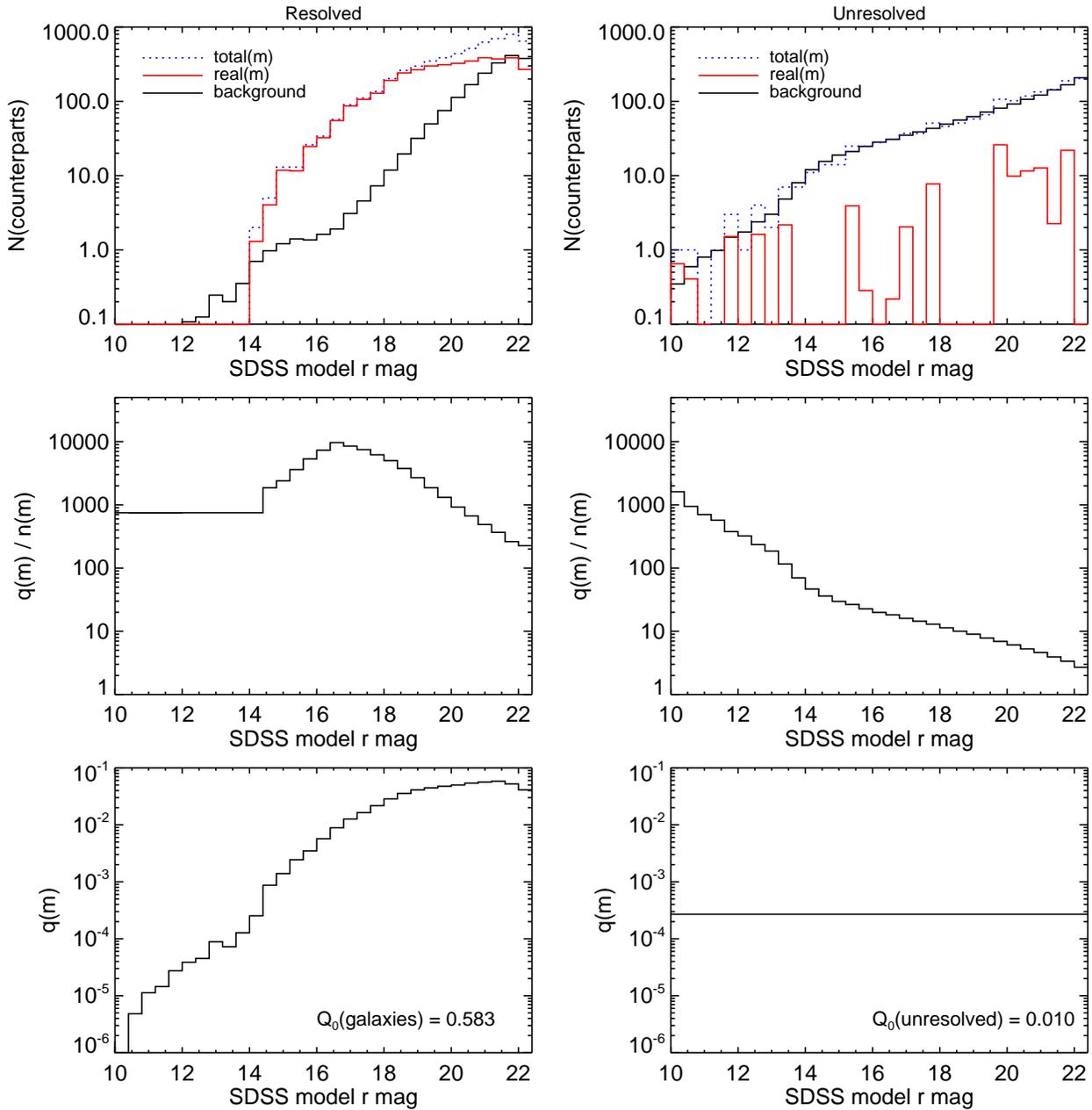}
\caption{Deriving the magnitude dependence of the LR for the resolved
  and unresolved counterparts to the 250$\mu$m catalogue (left and
  right columns, respectively). The analysis for the {\bf resolved}
  sources is discussed first, while the alternative procedure followed
  for the unresolved sources is described subsequently. {\em \bf Top
    Left}: Total$(m)$ (blue, dotted) represents the SDSS DR7 $r$--band
  model magnitude distribution of all the resolved sources that lie
  within 10.0 arcsec of the SPIRE 250\,$\mu$m centres. The black
  histogram represents the number of galaxies that we would expect
  within these search radii due to the background SDSS number counts
  alone. The red histogram, dubbed real$(m)$ as per Ciliegi et
  al. (2003), is the difference between the two, i.e. the SDSS DR7
  $r$--band model magnitude distribution of the excess sources above
  the background. {\em \bf Middle Left:} The ratio of $q(m) \slash
  n(m)$ represents the magnitude dependence of the $LR$. To avoid
  having a zero probability of a given source being the real
  counterpart due to our limited statistics on real$(m)$ (and hence
  $q(m)$) at bright magnitudes, we use the ratio of $q(m) \slash n(m)$
  in the brightest well-sampled bin ($r_{\rm mag} = 14.2$) to define
  the values of $q(m)$ for resolved sources with SDSS $r$--band
  magnitude $\le 14.0$. {\em \bf Bottom Left:} The resulting $q(m)$
  distribution -- our best estimate of the probability that a true
  counterpart to a 250\,$\mu$m source has a magnitude $m$ -- using the
  $n(m)$ distribution to overcome the small number statistics at
  bright $r$--band magnitudes. To further reduce the effects of noise,
  we boxcar smooth the $q(m)$ distribution for resolved sources with a
  3 bin kernel. $Q_0$ for the resolved sources is determined to be
  0.583. For the {\bf unresolved} sources (right column), which have
  considerably fewer excess sources (reflected in the lower value of
  $Q_0^{\rm unres}$), the corresponding analysis is slightly
  different. We define the magnitude dependence of the LR -- $q(m)
  \slash n(m)$ -- assuming a constant $q(m)$ (bottom panel, right
  column) normalised to reflect $Q_0^{\rm unres} = 0.010$. This
  situation will improve with the higher quality statistics that the
  full {\em Herschel}--ATLAS catalogue will produce.}
\label{fig:nm}
\end{figure*}

The distributions of $q(m)$, and $n(m)$ (as well as the magnitude
dependence of the LR -- $q(m)$\slash $n(m)$) are shown in Figure
\ref{fig:nm}, in which the left and right columns show the values for
the resolved and unresolved sources, respectively. While the $q(m)$
distribution for galaxies is well--sampled at $r \gtsim 14$\, mag, we
assume that $q(m) \slash n(m)$ is constant for all sources brighter
than this, enabling us to use our well-defined $n(m)$ to estimate
$q(m)$ for the brightest galaxies.

Since the fraction of {\em Herschel}--ATLAS sources associated with
unresolved counterparts is low (reflected in $Q_0^{\rm unres} =
0.010$), the method used to determine $q(m)$ for these sources
differs. In order to ensure that the LR results for stars\slash QSOs
are not dominated by small number statistics, we assume a flat prior
on $q(m)$, normalised to retain $Q_0^{\rm unres} = 0.010$ (figure
\ref{fig:nm}).

We can correct our value for $Q_0$ for the clustering of SDSS sources
by simply dividing $Q_0$ by $1 + \int_{0}^{10 {\rm arcsec}} w(\theta)
d\theta = 1.0008$ (remembering that $\theta$ is measured in degrees),
giving a clustering--corrected value of $\tilde Q_0 = 0.592$. This
value is broadly consistent with the recent results of Dunlop et
al. (2010), who recover optical counterparts to 8 out of 20
250\,$\mu$m sources brighter than 36\,$mJy$ in data from the BLAST
observations of the GOODS-South field to a comparable $i$--band
magnitude (albeit with lower angular resolution at 250\,$\mu$m and
much more sensitive optical, infrared and radio data), while Dye et
al. (2009) found 80 counterparts to the 175 BLAST 250$\mu$m sources
brighter than 55\,$mJy$ down to similar magnitude limits in $r$-- or
$R$--band data (S. Dye, private communication).

\begin{figure}
\centering
\includegraphics[width=0.99\columnwidth]{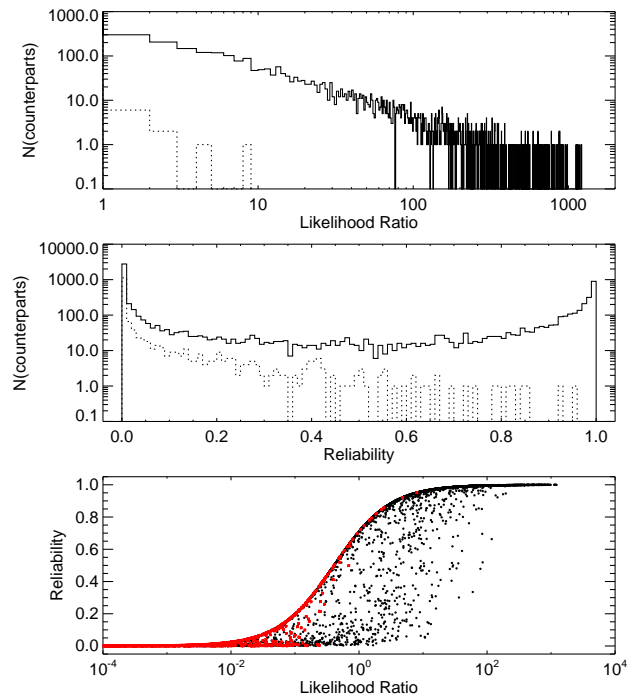}
\caption{Top: Histogram of the Likelihood Ratio values for all of the
  7230 potential counterparts to the 6621 5$\sigma$ sources in the
  SPIRE SDP catalogue. The LR values for the resolved sources
  (i.e. galaxies) are shown as the solid histogram, with unresolved
  sources shown as the dotted histogram. Middle: Reliabilities for
  each counterpart. Once more, the solid histogram represents the
  resolved sources, while the dotted histogram represents the
  unresolved sources. There are a total of 2423 sources which have a
  reliability $\ge 0.8$, of which five are unresolved using the
  star\slash galaxy separation criteria of Baldry et
  al. (2010). Bottom: The variation of the reliability as a function
  of the likelihood ratio. This is not a linear relation since some
  sources have more than one counterpart with a high likelihood
  ratio. There are 263 SDSS $r$--band sources with reliability $< 0.8$
  but $L > 1.63$ (the value above which $R > 0.8$ for a single
  counterpart within the 10.0 arcsec maximum search radius). These may
  be interacting systems, as discussed in section \ref{mergers}. These
  sources also demonstrate a possible limitation of the LR method,
  since the method implicitly assumes that there is only one true
  counterpart to a given 250\,$\mu$m source.} \label{fig:LRhist}
\end{figure}

To account for the fact that an {\em Herschel}--ATLAS source may have
more than one possible counterpart, we also define a reliability $R_j$
for each object $j$ being the correct counterpart out of all those
counterparts within $r_{\rm max}$, again following Sutherland \&
Saunders (1992):

\begin{equation}
R_j = \frac{L_j}{\sum_iL_i + (1 - Q_0)},
\label{eqn:rel}
\end{equation}

\noindent where the LR values have been determined for the resolved
and unresolved counterparts separately (see Figure
\ref{fig:LRhist}). The reliability is a key statistic; we recommend
using only those counterparts with reliability $R \ge 0.8$ for
analysis, since this ensures not only that the contamination rate is
low (see below), but also that only one $r$--band source dominates the
far--infrared emission (as required for e.g. deriving spectral energy
distributions for 250\,$\mu$m--selected galaxies in the {\em
  Herschel}--ATLAS catalogue, Smith et al. {\it in prep}). This is
more conservative than other works in the literature (e.g. Chapin et
al., 2010), where the chosen LR limit was defined based on a 10
percent sample contamination rate.

In order to estimate the number of false IDs in our reliable sample,
we calculate:

\begin{equation}
N(\mathrm{false}) = \sum_{R \ge 0.8}(1-R).
\end{equation}

\noindent As a result we expect 103 false IDs in our sample, which
corresponds to a contamination rate of 4.2\%. For those investigations
in which it is desirable only to determine whether an optical source
is associated with an {\em Herschel}--ATLAS object (with additional
caveats about lensed sources and the de-blending efficiency in the
optical catalogue), it is sufficient to use a likelihood ratio cut
(e.g. $L > 5.0$, i.e. the source is 5 times more likely to be
associated with the sub--millimetre object than it is to be a chance
superposition of sources). This aspect of the likelihood ratio
technique is discussed in more detail in section \ref{mergers}.

In Table \ref{tab:nmatch}, we present the number of possible optical
counterparts within 10.0 arcsec of the 250\,$\mu$m sample, including
the relative fractions of reliable associations. Only half of the
250\,$\mu$m sources with a single optical counterpart within the
search radius are deemed reliable.

\begin{table}
\caption{The distribution of the number of SDSS $r$--band sources
  within 10.0 arcsec of the 250\,$\mu$m positions, and the fraction of
  reliable counterparts. There are 2869 sources with only one possible
  match within 10.0 arcsec, and yet only 1389 of these are determined
  to be reliable; the vast superiority of the LR technique over a
  simple nearest-neighbour algorithm is evident.}  \centering
  \begin{tabular}{|c|c|c|c|}
    \hline
    $N$(matches) & $N$(250\,$\mu$m sources) & $N$($R \ge 0.8$) & \% \\
    \hline
    0 & 1865 &  & \\
    1 & 2869 & 1389 & 48.4 \\
    2 & 1400 & 782 & 55.9 \\
    3 & 400 & 210 & 52.5 \\
    4 & 76 & 38 & 50.0 \\
    5 & 9 & 4 & 44.4 \\ 
    6 & 2 & 0 & 0.0 \\
    \hline
    TOTAL: & 6621 & 2423 & 36.6 \\
    \hline
    \label{tab:nmatch}
  \end{tabular}
\end{table}

To estimate the fraction of 250\,$\mu$m sources with a counterpart
above our detection limit recovered as having $R \ge 0.8$, we assume
that 250\,$\mu$m--selected SDSS sources cluster in the same way as
SDSS $r$--band--selected sources (the results of Maddox et al., 2010,
suggest that this assumption is reasonable). Under this assumption, we
may calculate the completeness, $\eta$, of the reliable sources in our
sample:

\begin{equation}
\eta = \frac{n(R \ge 0.8)}{n(250\mu m > 5\sigma)}\times \frac{1}{\tilde
  Q_0}.
\label{incompleteness}
\end{equation}

\noindent We have reliably identified $\eta = $61.8 percent of the
optical counterparts bright enough to be detected in the SDSS
$r$--band catalogue. This constitutes an overall identification rate
of 36.6 percent for $\ge 5\sigma$ 250\,$\mu$m sources in the {\em
  Herschel}--ATLAS SDP observations.

\subsection{Checking the identification process}

By selecting sources at 250\,$\mu$m rather than longer wavelengths,
the negative $k$-correction that results in e.g. 850\,$\mu$m--selected
galaxies residing at a median redshift of $z \sim$2 (Chapman et al.,
2003) has a much less dramatic effect, and observations have shown
that a significant fraction of 250\,$\mu$m--selected galaxies reside
at $z < 1$ (e.g. Chapin et al., 2010, Dunlop et al., 2010, Dye et al.,
2010). As a result, their optical counterparts will be much brighter
than 850\,$\mu$m--selected sources, and therefore readily detectable
by shallower optical\slash near--infrared imaging with a much lower
source density. For the SDSS DR7 $r$--band source catalogue that we
use for the purposes of this investigation, we expect only $\sim0.48$
background sources within the 10.0 arcsec search radius, down to the
magnitude limit of $r = 22.4$\,mag (of these, $\sim$\,0.26\slash 0.22
will be resolved\slash unresolved, respectively). Furthermore, these
background sources may be expected to be evenly distributed throughout
the area within the maximum search radius, unlike the true
counterparts.

We performed the following simple checks to determine the
effectiveness of the LR technique for the {\em Herschel}--ATLAS SDP
catalogue.

\subsubsection{LR analyses of random catalogues}
\label{sec:randcat}

\begin{figure}
\centering
\includegraphics[width=0.95\columnwidth]{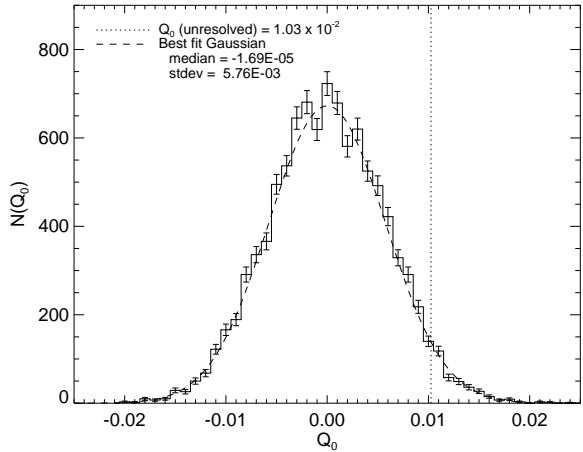}
\caption{Histogram of the 10000 realisations of determining $Q_0$ for
  6621 randomly--positioned {\em Herschel}--ATLAS sources, designed to
  determine the 1$\sigma$ uncertainty on $Q_0$. A Gaussian
  distribution with median and standard deviation derived from the
  histogram is overlaid (dashed line), and the value of $Q_0^{\rm
    unres}$ is indicated by the vertical dotted line.}
\label{fig:Q0}
\end{figure}

As a first test of whether the LR technique produces sensible results,
we wanted to test the method in the absence of any true association
between the 250\,$\mu$m and SDSS positions. We randomised the
positions of the 6621 sources and re-ran the LR analysis 10000 times,
recording the derived value of $Q_0$ each time. The histogram of the
resulting $Q_0$ distribution had a median of 0.000 with a $1\sigma$
uncertainty of $0.006$. In these cases, where $Q_0 \approx 0$, the
values for $L$ and hence $R$ are unreliable, since the distributions
of total$(m)$ and real$(m)$ that we determine are almost identical,
and the latter is strongly affected by noise as a result (see section
\ref{sec:lr_magdep}). The histogram of the simulated values for $Q_0$
is shown in Figure \ref{fig:Q0}, with a Gaussian distribution with
appropriate median and standard deviation overlaid (dashed line). The
derived value of $Q_0^{\rm unres}$ determined in section
\ref{sec:lr_magdep} is overlaid as the vertical dotted line. With
$Q_0^{\rm unres}$ residing within $2\sigma$ of the median $Q_0$ value
for these random catalogues, it is not clear that the population of
unresolved sources is detected in the {\em Herschel}--ATLAS SDP data
(see section \ref{unresolved}). However, to avoid the possibility of
missing real counterparts of potentially great scientific importance,
we must not ignore the possibility that unresolved sources are
detected in our 250\,$\mu$m catalogue.

\subsubsection{LR analyses of SDSS galaxies}

We also performed a test in which we replaced the SPIRE positions in
our LR analysis with the positions of SDSS galaxies, while retaining
the $250\mu$m fluxes and errors in order to accurately reproduce the
positional uncertainties according to the method given in section
\ref{sec:f_r}. We found that we recovered 97.2\%\ of the SDSS galaxies
at $R \ge 0.8$, which reduced to 93.9\%\ when the SDSS positions were
varied according to a Gaussian positional offset with standard
deviation appropriate for the signal-to-noise ratio of the real SPIRE
sources (according to the rescaled formula given in section
\ref{sec:f_r}). Comparing this value to our overall ID rate we see
that we are not typically missing counterparts because we have
underestimated their positional errors, but because of the fact that
approximately 39\%\ of 250\,$\mu$m sources (the actual value is
$1-\tilde Q_0$) are not detected in SDSS $r$--band data down to the
magnitude limit of our survey.

We also compared our new catalogues with the small subsets of
overlapping objects in the Imperial {\it IRAS}-FSC Redshift Catalogue
(IIFSC$z$, Wang \&\ Rowan-Robinson, 2009) and FIRST surveys (Becker,
White \& Helfand, 1995). These comparisons are presented in detail in
appendix \ref{sec:checks}, but to summarize:

\begin{itemize}
\item We find updated positions for five IIFSC$z$ galaxies which
  previously had misidentified or unidentified counterparts. 
\item Our catalogue is consistent with a catalogue of FIRST radio
  sources matched to the 250\,$\mu$m sample, provided that the radio
  sources are detected in our optical data, and that the optical
  counterparts contain only single components at moderate separations
  from the 250\,$\mu$m centroids.
\item A small collection of {\it Spitzer Space Telescope} snapshot
  images taken at near infrared wavelengths reinforce our belief in
  the accuracy of our method.
\end{itemize}

\section{Redshifts in the {\em Herschel}--ATLAS 9hr field}
\label{sec:redshifts}

\subsection{Spectroscopic Redshifts}

The GAMA catalogue (Driver et al., 2010) contains 12,626 new
spectroscopic redshifts in the {\em Herschel}--ATLAS SDP region for
sources satisfying the GAMA target selection criteria (including
magnitude limits of $r < 19.4$, $z < 18.2$ \& $K < 17.6$ -- Baldry et
al. 2010). In addition, there are a further 3281 redshifts available
in this region from the SDSS DR7, 248 from the 2SLAQ-LRG survey
(Cannon et al., 2006), 939 from the 2SLAQ-QSO survey (Croom et al.,
2009) and 29 from the 6dFGS (Jones et al., 2009). 1099 spectroscopic
redshifts for reliable counterparts were collated (including those
from the SDSS DR7, 6dFGS, 2SLAQ-QSO\slash LRG surveys and the GAMA
catalogue), meaning that 41.0\%\ of our $R \ge 0.8$ counterparts have
spectroscopic redshifts (and 15.0\%\ of all $\ge 5 \sigma$ 250\,$\mu$m
sources). We note that none of the spectroscopic catalogues (with the
exception of the 6dFGS catalogue) extends to declinations less than -1
deg. The number of redshifts for reliable counterparts from each
spectroscopic survey is presented in table \ref{tab:redshift_sources},
and the redshift properties of 250\,$\mu$m selected galaxies are
discussed in section \ref{sec:zdist}.

\begin{table*}
\caption{The number of spectroscopic redshifts for reliable
  counterparts from each survey used in our final catalogue. The
  majority of the spectroscopic redshifts used in our follow-up
  studies of the reliable 250\,$\mu$m associations come from
  GAMA. There are a total of 1099 $R \ge 0.80$ 250$\mu$m counterparts
  with spectroscopic redshifts.} \centering
\begin{tabular}{llccc}
\hline
\multirow{2}{*}{Survey} & \multirow{2}{*}{Reference} & \multicolumn{2}{c}{Number of redshifts} & Percentage of \\
%\cline{3-5}
  &  & $R \ge 0.8$ & Catalogue & {\em Herschel}--ATLAS IDs \\
\hline
2SLAQ-LRG & Cannon et al. (2006) & 3 & 248 & 0.12 \\
2SLAQ-QSO & Croom et al. (2009) & 4 & 939 & 0.17 \\
6dFGS & Jones et al. (2009) & 12 & 29 &  0.50  \\ 
GAMA & Driver et al. (2010) & 766 & 12626 &  31.6 \\
SDSS DR7 & Abazajian et al. (2009) & 316 & 3281 & 13.0 \\
\hline
\label{tab:redshift_sources}
\end{tabular}
\end{table*}

\subsection{Photometric Redshifts}
\label{sec:photoz}

For those sources without spectra, we estimate photometric redshifts
using optical and near--infrared photometry. The {\em Herschel}--ATLAS
SDP field has almost complete optical coverage in $ugriz$ from the
SDSS DR7 and near--infrared $YJHK$ photometry from 7th data release of
the UKIDSS Large Area Survey (Lawrence et al. 2007). As well as having
spectroscopy from GAMA and SDSS, these very wide-area surveys overlap
with several deeper spectroscopic surveys (Davis et al., 2003, Cannon
et al., 2006, Davis et al., 2007, Lilly et al., 2007) which allow us
to construct a spectroscopic training set with large numbers of
objects ($>1000$ per bin of unit magnitude or 0.1 in redshift) up to
$r$--band magnitudes $r<23$ and redshifts $z<1.0$, i.e. to
approximately the photometric depth of SDSS and UKIDSS-LAS.

This large and relatively complete training set allows us to use an
empirical regression method to estimate the photometric redshifts.  We
use the well-known artificial neural network code {\sc{annz}}
(Collister \& Lahav, 2004) with a network architecture of $N:2N:2N:1$,
where $N$ is the number of photometric bands used as inputs; although
there are 9 photometric bands ($ugrizYJHK$) available in this case, we
ignore bands where an object has no coverage or where the photometry
is flagged as dubious, and train separate neural networks for all
combinations of bands with at least three good detections.  An
advantage of {\sc{annz}} is that it provides redshift error estimates,
$\sigma_z$, based on the photometric errors it is supplied with; we
checked that these errors were distributed correctly by confirming
that, for a set of validation data with spectroscopic redshifts,
$(z_{\mathrm{phot}} - z_{\mathrm{spec}})/ \sigma_z$ follows a Gaussian
distribution centred on zero -- however we found that the width of the
best-fitting Gaussian was $\sim 1.4$, indicating that the errors were
underestimated by this factor on average.  To improve the accuracy of
the error estimates, we used the width of this distribution to correct
the error estimates individually for each trained network, with
correction factors of typically 1.3 to 2.

Confirming the accuracy of empirical photometric redshifts is always
difficult, since the objects for which we have spectroscopic redshifts
for comparison are, by necessity, drawn from the same sample which is
used to train the neural network, and may not be fully representative
of the whole population of objects to which the method is applied.  A
particular concern in our case was that the {\em Herschel}
counterparts are likely to have a different distribution in colour
space than the training-set galaxies, and so any bias in the
photometric redshift as a function of colour could cause the average
redshift of {\em Herschel} counterparts to be systematically wrong.
However, we satisfied ourselves that this was not a problem by looking
for trends in the difference between photometric and spectroscopic
redshift as a function of colour, and finding no significant trend
with any colour.  For example, the best-fitting straight line
relationship between ($z_{\mathrm{phot}}-z_{\mathrm{spec}}$) and
$(r-K)$, for objects in our validation dataset, has a gradient of
0.00035 -- two orders of magnitude smaller than the scatter of
($z_{\mathrm{phot}}-z_{\mathrm{spec}}$), which for the same sample has
a standard deviation of 0.037 overall.

In the event that the counterparts are so obscured that they are
invisible in the optical data, they will clearly have unreliable
photometric redshifts (this is inevitable, given the small number of
detections that would be available), but this scenario will become
apparent as large errors on the photometric redshifts of these
sources. In any case, these sources will not pass our $r < 22.4$\,mag
selection criterion. With this new catalogue of photometric redshifts,
all sources detected in at least three photometric bands have either a
spectroscopic or photometric redshift.

\subsection{Redshift distribution and completeness of reliably-identified 250\,$\mu$m sources}
\label{sec:zdist}

\begin{figure}
\centering
\includegraphics[width=0.99\columnwidth]{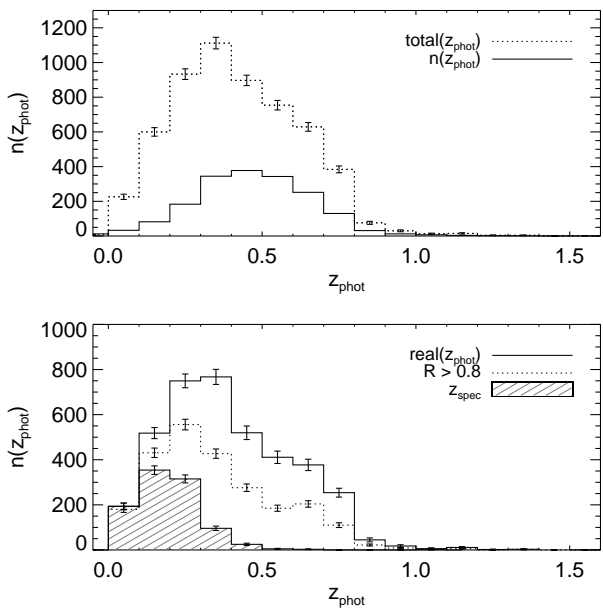}
\caption{Top: total$(z)$ (dotted histogram) represents the photometric
  redshift distribution of all of the sources within 10.0 arcsec of a
  250\,$\mu$m position, while $n(z)$ (solid line) represents the
  expected background based on the area covered and the redshift
  catalogue number counts. We note that the histogram of background
  sources is consistent with the results of Oyaizu et
  al. (2008). Bottom: real$(z)$ (solid line) shows the redshift
  distribution of the excess sources above the background around the
  250\,$\mu$m centres, and is compared with the dotted line which
  shows the photometric redshift distribution for those counterparts
  with $R \ge 0.8$ in our LR analysis. We note that real$(z)$ peaks at
  a lower redshift than the intrinsic $n(z)$ for SDSS galaxies (solid
  histogram, top panel). The shaded histogram shows the number of
  spectroscopic redshifts for galaxies (as defined by the star galaxy
  separation criteria in Baldry et al., 2010) in our sample. The
  percentage completeness in our catalogue is given in Table
  \ref{tab:z_completeness}.}
\label{fig:completeness}
\end{figure}

Using a method analogous to that used in section \ref{sec:lr_magdep}
to calculate the $r$--band magnitude distribution of counterparts to
250\,$\mu$m sources, we may determine the completeness of our
reliably--matched 250\,$\mu$m and SDSS objects as a function of
redshift. Here we define the completeness as the fraction of reliably
identified counterparts, compared with all of those counterparts that
are detected in our data i.e. $Q_0 \times 6621 \approx 3925$
sources. First, we use the star-galaxy separation method from Baldry
et al. (2010 -- discussed in section \ref{unresolved}) to ensure that
only galaxies remain in our sample.  We then use the catalogue of
photometric and spectroscopic redshifts discussed in section
\ref{sec:photoz} to determine the distribution of the galaxy redshifts
in the catalogue, $n(z)$. Since we know that this catalogue covers an
area of 36.0 deg$^2$, we can scale to the total sky area searched
around the 250\,$\mu$m sources ($6621 \times \pi r_{\rm max}^2$) to
determine the expected background, n$(z)$ (see Figure
\ref{fig:completeness}). Subtracting total$(z)$ (the $z$ distribution
of all sources within 10.0 arcsec of the 250\,$\mu$m centres) from
$n(z)$ then gives us the number of excess sources around the
250\,$\mu$m positions, real$(z)$, and by comparing this with the
photometric redshift distribution for those sources with $R \ge 0.8$
we can estimate the completeness of our reliable catalogue as a
function of $z$. The completeness values for our reliable catalogue in
bins between $0.0 < z < 1.1$ are presented in Table
\ref{tab:z_completeness}.

\begin{table}
\caption{The percentage completeness of our reliable catalogue as a
  function of photometric redshift. These values are derived as
  described in section \ref{sec:zdist}, using a method analogous to
  that applied to determine the $r$--band magnitude distribution of
  250\,$\mu$m sources. The errors are determined assuming that they are
  dominated by the Poissonian errors on real$(z_{\rm phot})$.}
\centering
  \begin{tabular}{|c|c|c|}
    \hline
    $z_{\rm phot}$ & Completeness (\%) & $\sigma_{\rm comp}$ \\
    \hline
    0.0--0.1 & 93.2 & 7.5 \\
    0.1--0.2 & 83.2 & 4.8 \\
    0.2--0.3 & 74.2 & 4.2 \\
    0.3--0.4 & 55.6 & 4.8 \\
    0.4--0.5 & 53.1 & 6.0 \\
    0.5--0.6 & 45.0 & 7.4 \\
    0.6--0.7 & 54.1 & 7.0 \\
    0.7--0.8 & 43.3 & 10.0 \\
    0.8--1.1 & 52.7 & 16.7 %\hline
    \label{tab:z_completeness}
  \end{tabular}
\end{table}

Figure \ref{fig:completeness} also shows the redshift distribution of
250\,$\mu$m--selected {\em Herschel}--ATLAS sources with reliable
counterparts and spectroscopic redshifts in our GAMA\slash SDSS DR7
catalogue. The redshift distribution of reliable $r$--band
counterparts peaks at $z_{\mathrm{phot}} = 0.25 \pm 0.05$, with a
median value of $z_{\mathrm{phot}} = 0.31^{+0.28}_{-0.15}$, or
$z_{\mathrm{spec}} = 0.18^{+0.10}_{-0.09}$ if only spectroscopic
redshifts are included (here the errors on the peak are based on the
half-width of one histogram bin, and those on the median values are
derived according to the 16th and 86th percentiles of the redshift
cumulative frequency distribution). The disparity between these median
redshift values is to be expected since the photometric redshifts are
computed to fainter magnitude limits than the spectroscopic redshifts
have been measured in the GAMA 9hr catalogue.

\begin{figure}
\centering
\includegraphics[width=0.99\columnwidth]{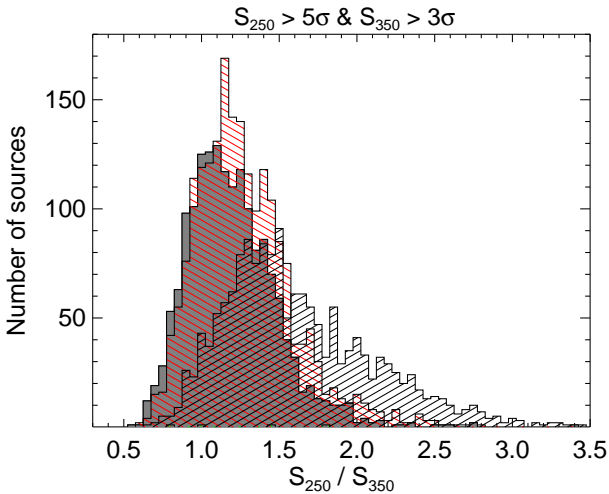}
\caption{Histograms of the SPIRE colours for those sources with
  reliable galaxy counterparts ($R \ge 0.8$, shaded black), those for
  which the most reliable candidate has $R<0.8$ (shaded red), and
  those remaining sources in the MAD-X catalogue without any SDSS
  $r$--band counterparts. All sources are detected at $\ge 5 \sigma$
  in the 250\,$\mu$m band, and $\ge 3\sigma$ in the 350\,$\mu$m
  band. Those galaxies for which we identify $R \ge 0.8$ counterparts
  have considerably bluer average $S_{250}/S_{350}$ colours than those
  for which we are unable to identify a counterpart. }
\label{fig:colours_reliability}
\end{figure}

It is also interesting to note that the photometric redshift
distribution of excess 250\,$\mu$m sources -- real$(z_{\rm phot})$ --
peaks at a lower redshift than the intrinsic redshift distribution of
the SDSS photometric redshift catalogue, $n(z_{\rm phot})$. This is
not due to our inability to reliably identify sources at higher
redshifts, since the N$(z)$ includes statistical non-detections, and
indeed the magnitude distribution of the true counterparts (calculated
in section \ref{sec:lr_magdep}) peaks at brighter magnitudes then the
background $n(m)$. This determination of real$(z_{\rm phot})$
indicates that the low redshift population of {\em Herschel}--ATLAS
galaxies in our 5$\sigma$ sample is generally at lower redshift than
the average SDSS galaxy, raising an interesting question about the
$\sim40$\% of {\em Herschel}--ATLAS sources which do not have a
counterpart above the SDSS DR7 limit. The redder sub-mm colours of
these blank field sources suggests that they are at much higher
redshifts (see Figure \ref{fig:colours_reliability}, and Section
\ref{submmcolours}), and furthermore, the study of {\em
  Herschel}--ATLAS colours by Amblard et al. (2010) indicates a second
population of sources at $z\sim 2$. The fact that we do not see a
rising $n(z)$ for {\em Herschel}--ATLAS sources in SDSS out to the
SDSS limit suggests that the total {\em Herschel}--ATLAS $n(z)$ is
bimodal, with a low-redshift peak at $z \approx 0.35 \pm 0.05$ (where
the error is once more derived according to the width of one histogram
bin) and a higher-redshift peak at $z>1$. Such behaviour is predicted
in several models of sub-mm galaxy populations (e.g. Lagache et
al. 2004; Negrello et al. 2007, Wilman et al., 2010) and also
suggested by stronger clustering in samples of {\em Herschel}--ATLAS
galaxies selected to have redder far--infrared colours (Maddox et
al. 2010), as well as the steep upturn in the {\em Herschel}--ATLAS
number counts at fluxes below 100 mJy (Clements et al. 2010) and the
results of BLAST, which include deeper optical samples with fainter
spectroscopy, albeit with smaller object samples by more than an order
of magnitude (Dunlop et al. 2010 and Chapin et al. 2010).

\section{Results}
\label{results}

The resulting values for the likelihood ratio and reliability for the
6621 5$\sigma$ sources in the 250\,$\mu$m selected catalogue are shown
in Figure \ref{fig:LRhist}. Approximately 58\%\ of galaxies with
$S_{\mathrm{250\,\mu m}} \ge 32$\,mJy are detected in our $r \le
22.4$\,mag SDSS catalog, and of those we identify 2423 counterparts
with reliability $R \ge 0.8$, which we consider robust. Of these
reliable counterparts, 1252 also have GALEX detections in at least one
ultraviolet band, and each source has either a reliable spectroscopic
redshift (1099 galaxies) or photometric redshift.

\begin{figure*}
\centering
\subfigure{\includegraphics[width=0.99\columnwidth]{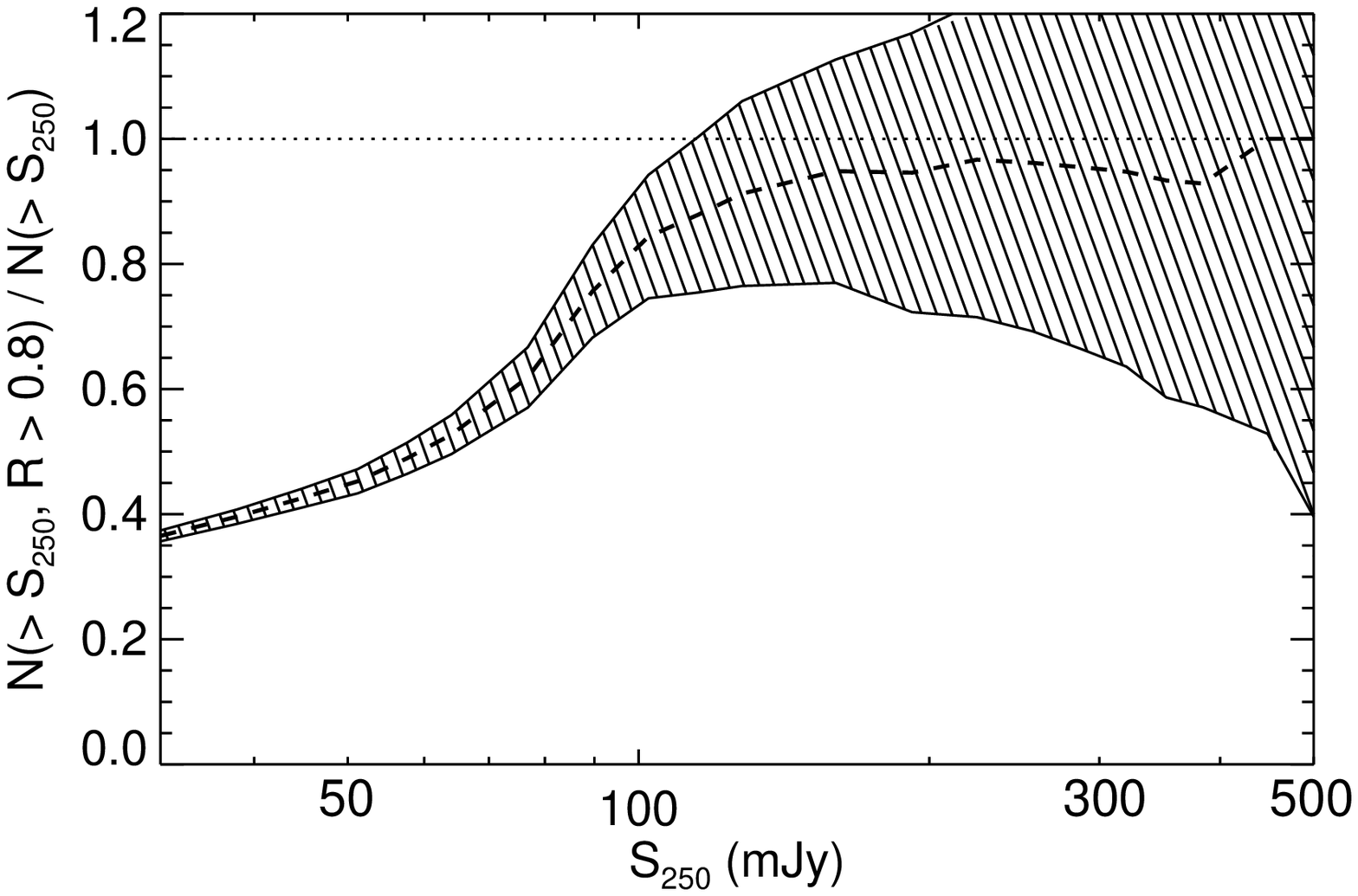}}
\subfigure{\includegraphics[width=0.99\columnwidth]{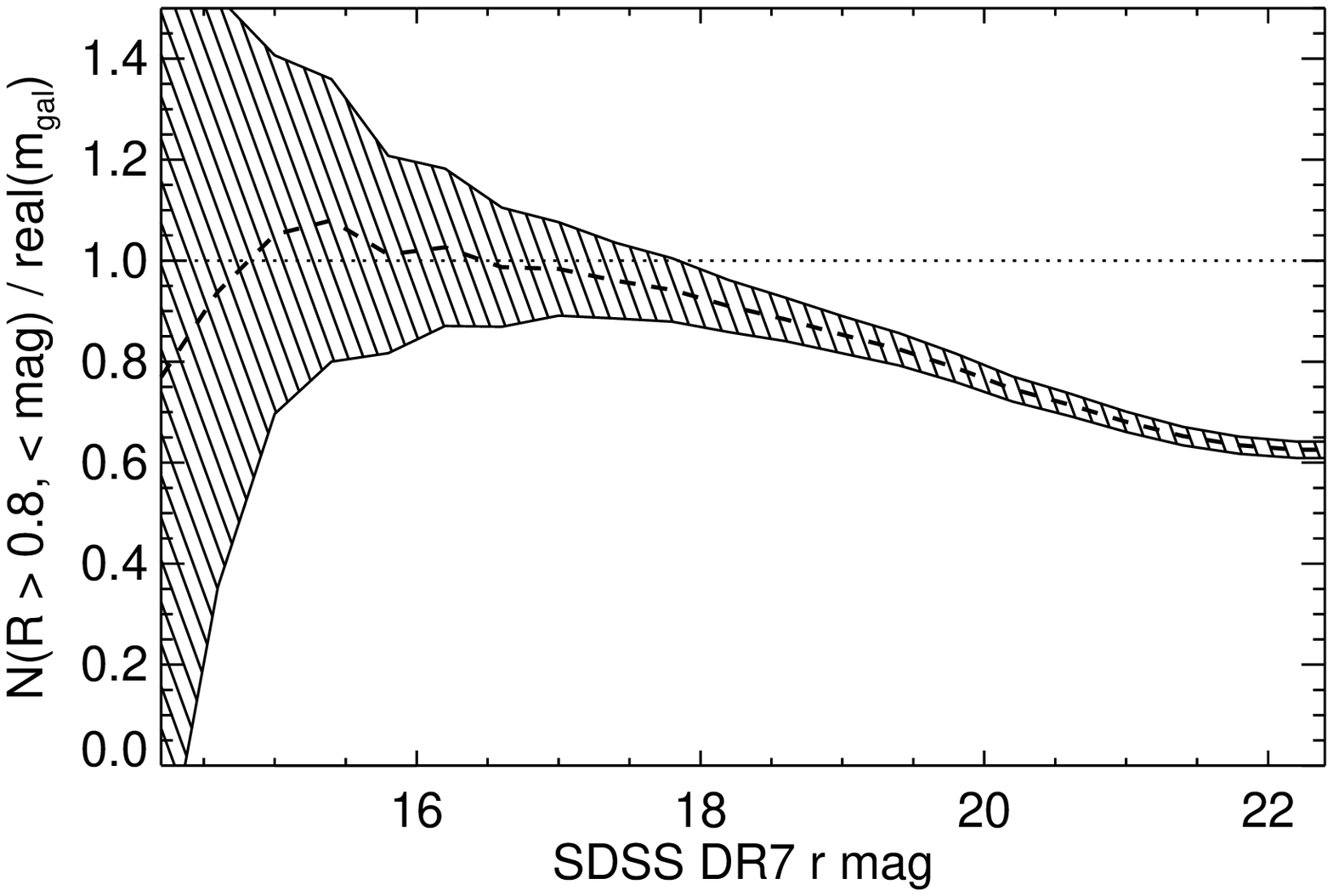}}
\caption{Left: Completeness of our identification catalogue as a
  function of 250\,$\mu$m flux. The ordinate indicates the fraction of
  250\,$\mu$m sources brighter than the flux given by the abscissa,
  for which we have reliably identified counterparts in the SDSS DR7
  $r$--band data. We can reliably identify the counterparts to 36.6
  percent of the 6621 sources in our 5$\sigma$ 250\,$\mu$m--selected
  catalogue down to the limit of our $r$--band SDSS DR7 catalogue.
  Right: The fraction of sources with statistical identifications in
  our SDSS DR7 $r$--band catalogue (i.e. real$(m)$) which can be
  reliably identified with $R > 0.80$, not accounting for the value of
  $Q_0$. We reliably identify $\sim$\,63\% of those counterparts that
  are detected in our $r < 22.4$\,mag survey data. The shaded regions
  indicate the 1$\sigma$ uncertainties in these completeness
  fractions, determined based on the Poisson errors on the number
  counts.}
\label{fig:idfrac}
\end{figure*}

Figure \ref{fig:idfrac} shows the fractional completeness in our
identification catalogue as a function of the 250\,$\mu$m flux, and of
the SDSS DR7 $r$--band magnitude of the counterparts. The shaded areas
indicate the 1$\sigma$ uncertainty on the completeness derived from
the Poisson errors on the number of sources brighter than a given
magnitude\slash flux.

\subsection{The sub-millimetre colours of SPIRE sources}
\label{submmcolours}

In Figure \ref{fig:colours_reliability} we display the $S_{250} /
S_{350}$ colours of the 250\,$\mu$m sources from the {\em
  Herschel}--ATLAS SDP catalogue with detections at $\ge 5 \sigma$ in
the 250\,$\mu$m band, and $\ge 3 \sigma$ in the 350\,$\mu$m band. The
sources have been divided into three sub-sets; those sources with
reliable $r$--band counterparts classified as galaxies ($R \ge 0.8$,
black shaded histogram), those for which the most reliable candidate
has $R < 0.8$ (red shaded histogram) and all of the remaining sources
in the MAD-X catalogue without any $r$--band counterparts (grey shaded
histogram). The median SPIRE colours for the three samples are quite
different, with median colours of $S_{250} \slash S_{350} = 1.51,
1.23, \&\ 1.16$ for the three respective samples. It is clear that the
$R \ge 0.8$ sources are considerably bluer than the SPIRE sources
without (reliable) counterparts, and a series of Kolmogorov--Smirnov
tests confirms that no two of the three sets of histograms are drawn
from the same parent distribution at $\gg99.9999$\% confidence.

The differences between the colour populations may be due to those
sources without reliable counterparts residing at higher redshift than
those for which we can identify reliable counterparts, causing the
peak of each such source's far--infrared spectral energy distribution
to move to longer wavelengths.

% We also note that those
%%four 
%three unresolved
%sources satisfying the signal--to--noise criteria demonstrate
%considerably redder colours than galaxies residing at redshifts $z
%\sim 0.0$ (which demonstrate the bluest colours in Figure
%\ref{fig:colours_reliability}). The bluest two of these 
%%four 
%three sources are spectroscopically confirmed SDSS QSOs at $z = 1.02$
%\&\ $1.87$, but the 
%%others have $S_{250}\slash S_{350} = 1.81 \pm
%%0.37$ \&\ $1.44 \pm 0.24$, suggesting that the dust in these possibly
%%stellar systems 
%other has $S_{250}\slash S_{350} = 1.81 \pm 0.37$, suggesting that the
%dust in this possibly stellar system may be considerably colder than
%in low--$z$ galaxies, although the SEDs of such systems differ from
%those of galaxies at the same temperature. This source is one of the
%candidate debris disk systems discussed in Thompson et al., 2010.

It is also clear that those sources for which the most reliable
candidate counterpart has $R<0.8$ are a different population from
those for which we identify no $r$--band counterparts. Half of these
sources can be explained by the expected number which are above the
SDSS limit but for which we cannot determine a reliable counterpart,
and the other half simply have an unrelated SDSS $r$--band source
within the search radius. This is reflected in the histogram for these
sources having colours intermediate between the reliable counterparts
and the ``no potential counterparts'' samples - it contains roughly
equal fractions of both types of object (presumably high and low
redshift).

\begin{figure}
\centering
\includegraphics[width=0.99\columnwidth]{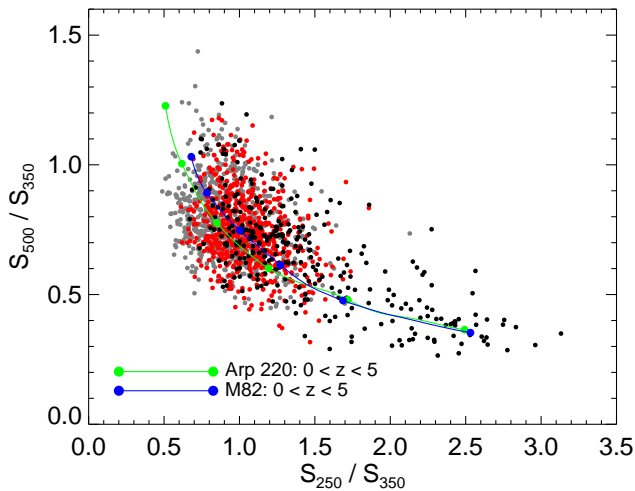}
\caption{SPIRE colour--colour diagram, showing the colours of
  5$\sigma$ 250\,$\mu$m and 350\,$\mu$m sources with $\ge 3\sigma$
  detections in the 500\,$\mu$m band. The colour scheme is as in
  Figure \ref{fig:colours_reliability}. The green and blue lines
  represent redshift colour tracks between $0.0 < z < 5.0$ based on
  Arp220 and M82 template SEDs from Silva et al., 1998, with solid
  circles along the tracks indicating the locations in colour--colour
  space of integer redshifts between these two values for that
  template.  }
\label{fig:spirecolours}
\end{figure}

We also compare the far--infrared colours with the results of Amblard
et al. (2010 -- their Figure 1 and our Figure \ref{fig:spirecolours}).
In this figure (which uses the same colour scheme as Figure
\ref{fig:colours_reliability}, with the $R \ge 0.8$ counterparts in
black) we consider only those sources at $\ge 5\sigma$ in the
250\,$\mu$m and 350\,$\mu$m bands and $\ge 3\sigma$ in the 500\,$\mu$m
bands to ensure a fair comparison. We identify 133 such sources with
$R \ge 0.8$ counterparts in our SDSS $r$--band catalogue.

\subsection{Lensed sources in {\it H}-ATLAS}

Wide--field sub--millimetre wavelength surveys such as the {\em
  Herschel}--ATLAS are particularly well-suited to detecting large
numbers of strongly--lensed sources (e.g. Blain, 1996, Negrello et
al., 2007), in which intrinsically faint distant galaxies may be
magnified by an otherwise unrelated foreground massive object along
the line of sight (e.g. galaxy, galaxy cluster), and observed at more
readily--detectable flux densities. Strong lensing can not only
amplify the brightness of these distant soures, but also increase
their angular size, allowing galaxies to be studied on scales smaller
than would otherwise be possible, making samples of strongly--lensed
galaxies an important cosmological probe (e.g. Swinbank et
al. 2010). At bright 500\,$\mu$m flux densities ($> 100$\,mJy), after
removing very local galaxies and blazars from the source counts, the
surface density of sources on the sky is dominated by strongly-lensed
galaxies, with large--area surveys such as {\em Herschel}--ATLAS
required to detect them due to their paucity on the sky ($\sim
0.5$\,deg$^{-2}$). This method of selecting lensed sources has one
huge advantage over other methods, in that the selection efficiency is
almost 100 percent (Negrello et al., 2010).

Of particular interest in Figure \ref{fig:spirecolours} is the
positioning of the 58 $R \ge 0.8$ galaxies with $S_{250} / S_{350} \le
1.5$. The redshift loci of Arp220 and M82 templates from Silva et
al. (1998), shown in green and blue respectively, suggest that sources
with such colours reside at $z > 1.0$, despite the photometric and
spectroscopic redshifts of their reliable optical counterparts
residing at $z < 1.0$ (Figure \ref{fig:completeness}). This disparity
is, for some of these sources, caused by the blending of galaxies in
the 350\slash 500\,$\mu$m bands (the results of Rigby et al., 2010,
suggest that the extracted 500\,$\mu$m flux densities of more than a
quarter of $>5\sigma$ sources are enhanced by factors of up to $\sim$2
due to multiple sources residing within a beam, for example).
However, it is also possible that some of these are intrinsically
high--redshift far--infrared sources which are strongly lensed by
low--redshift foreground galaxies. The models of Negrello et
al. (2007) predict that the fraction of lensed sources to these
sensitivity limits is $\sim$4\%. In our catalogue, $\sim$14\%\ of {\em
  Herschel}--ATLAS sources with low--redshift $R \ge 0.8$ counterparts
have $S_{250} / S_{350} \le 1.5$, consistent with high-$z$ galaxies
(209 out of 1480 sources that are detected at $\ge 5\sigma$ at 250 and
350\,$\mu$m and $\ge 3\sigma$ at 500\,$\mu$m). These numbers suggest
that approximately one third of these sources may be strongly-lensed
galaxies, although more realistic simulations will be required to
thoroughly test this interesting hypothesis.

\subsection{Multiple sources} 
\label{mergers}

\begin{figure*}
\centering
%\subfigure{\includegraphics[width=0.18\textwidth]{merger1422.eps}}
\subfigure{\includegraphics[width=0.46\textwidth]{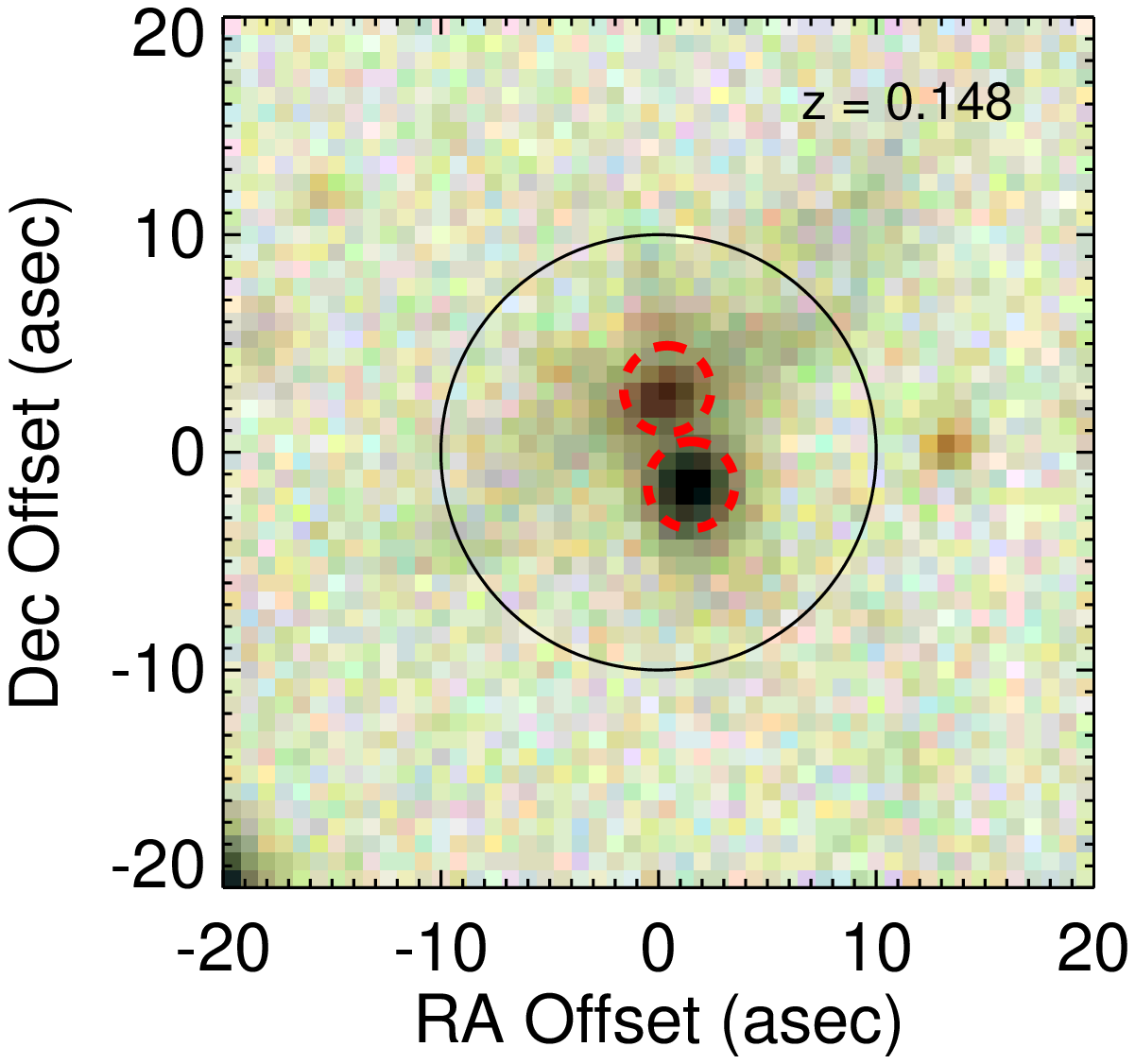}}
%\subfigure{\includegraphics[width=0.30\textwidth]{merger1133.eps}}
\subfigure{\includegraphics[width=0.46\textwidth]{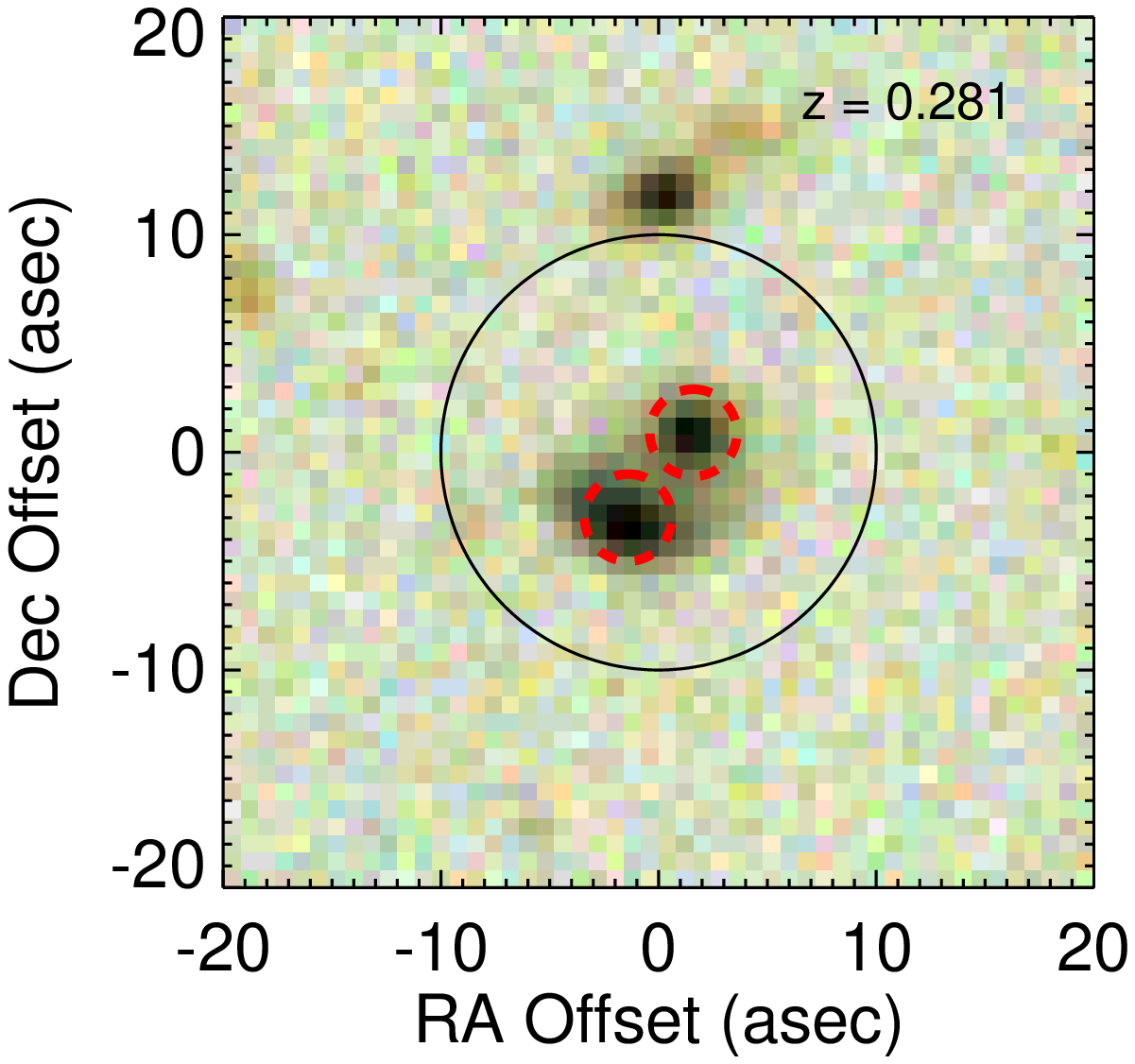}}
\subfigure{\includegraphics[width=0.46\textwidth]{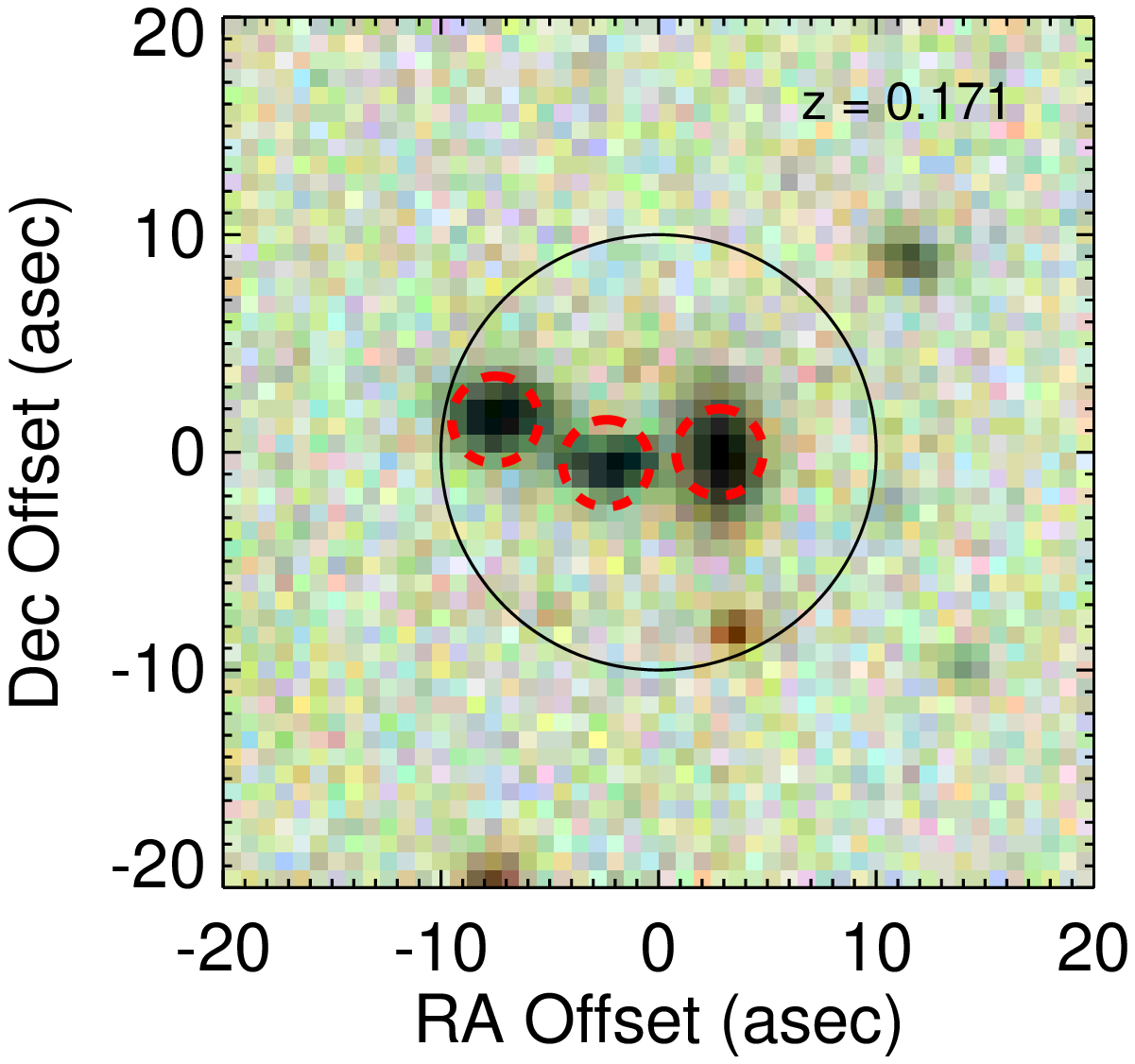}}
%\subfigure{\includegraphics[width=0.30\textwidth]{merger3650.eps}}
\subfigure{\includegraphics[width=0.46\textwidth]{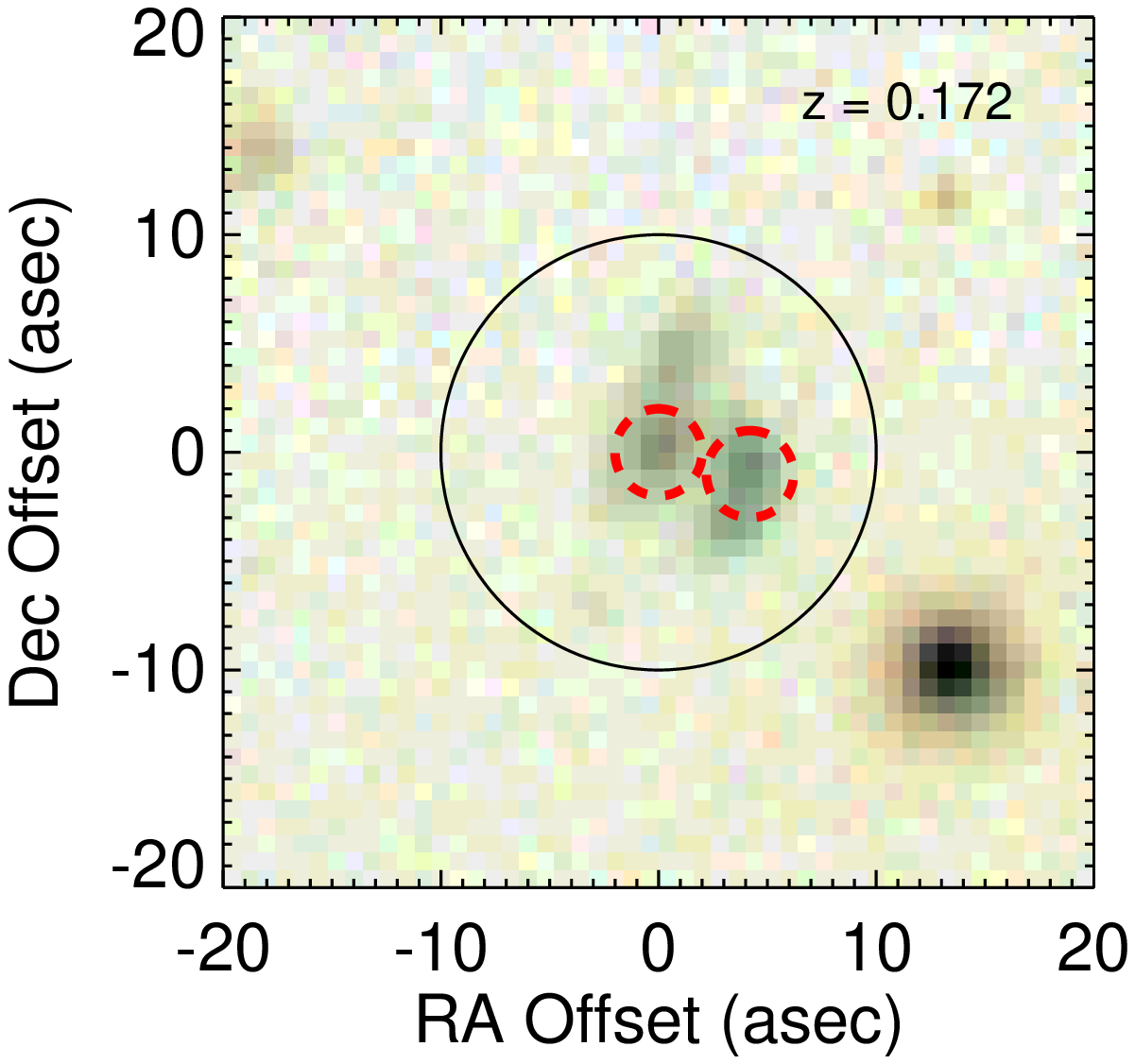}}
%\subfigure{\includegraphics[width=0.18\textwidth]{merger3650.eps}}
\caption{SDSS $gri$ colour images centred on the positions of four
  SPIRE 250\,$\mu$m 5$\sigma$ sources with at least one counterpart
  with $L > 5.0$ and $R < 0.8$, and with at least two spectroscopic
  redshifts within $\Delta z = 0.001$ of each other (indicated by the
  red dashed circles). Each image is 40 arcsec on a side orientated
  such that North is up and East is to the left.}
\label{fig:mergers}
\end{figure*}

We may also use our catalogue to identify 250\,$\mu$m sources with
multiple counterpart galaxies, by considering those which have at
least one $L > 5.0$, $R < 0.80$ optical source within 10 arcsec (of
course, this will also select low-probability superpositions of
sources on the sky, e.g. Arp, 1967). There are a total of 118 such
250\,$\mu$m sources which have at least one counterpart with $L > 5.0$
and $R < 0.8$ in our catalogue. It is possible that these sources
contain multiple interacting counterparts, and indeed four of these
sources have at least two counterparts with spectroscopic redshifts
with $\Delta z\, \ltsim 0.001$ (including one of the radio sources
mentioned in appendix \ref{radiocomp}, H--ATLAS
J090631.3+004605). SDSS three-colour images of each
spectroscopically-confirmed galaxy interaction are shown in Figure
\ref{fig:mergers}. There may be further examples for which we do not
have spectroscopic redshifts.

For a more ``complete'' sample of cross--identifications, sources
above some threshold in $L$ could be considered, however in this case
there is no immediate information to decide which of the multiple
counterparts contributes most to the SPIRE flux, without resorting to
priors on e.g. the colours of sources (e.g. Roseboom et al., 2009).
This is work which we are pursuing and will look to implement in our
next data release.

Finally, there is one additional source (H--ATLAS 090130.2-00215) with
two high LR counterparts that have differing spectroscopic redshifts.
This 250\,$\mu$m source has counterparts with $L = 15.8$ \& 35.0,
residing at $z_{\rm spec} = 0.196$ and $z_{\rm spec} = 0.255$,
respectively. The latter counterpart is also a $P<0.20$ radio source,
mentioned in appendix \ref{radiocomp}, and presumably constitutes one
of the low-probability superpositions mentioned above. We also note
that merging sources may have real positional offsets between the dust
emission in the far--infrared and the starlight which dominates the
optical (see e.g. Zhu et al., 2007, Ivison et al., 2008, or Smith et
al., 2010a).

\subsection{Unresolved sources}
\label{unresolved}

Although the main focus of this paper is the reliable identification
of galaxies selected at 250\,$\mu$m, we have also applied the LR
method separately to identify any reliable unresolved sources. We have
used our spectroscopic data set to further split the population of
unresolved sources in to groups of candidate stars and QSOs. There are
a total of five $R \ge 0.80$ unresolved sources in the 250\,$\mu$m
selected sample (green asterisks in Figure \ref{fig:stargalsep}), of
which three occupy the stellar colour-colour locus, while two have
colours or spectroscopic redshifts consistent with being QSOs.

Studying these objects in detail is beyond the scope of this paper,
but see e.g. Thompson et al., (2010) for a discussion of stellar
sources in the {\em Herschel}--ATLAS and the identification of two
candidate debris disks.

\section{Conclusions}
\label{conclusions}

We have demonstrated that the likelihood ratio method of Sutherland \&
Saunders (1992) is an appropriate way to determine reliable
counterparts for 250\,$\mu$m--selected galaxies from the {\em
  Herschel}--ATLAS science demonstration phase observations in the
SDSS DR7 $r$--band observations of the GAMA 9 hour field. We have
determined reliable ($R \ge 0.8$) counterparts to 2423 out of 6621
sources detected at a SNR $\ge 5$, and found that $\sim$59.3\%\ have
counterparts brighter than r = 22.4 (the limit of our catalogue). We
identify reliable counterparts to 36.6\%\ of the $250\mu$m sources
(2423 out of 6621), and our calculations in section
\ref{sec:lr_magdep} suggest that our sample is 61.8\%\ complete down
to the SDSS $r$--band limit of our catalogue, in the sense that we
have reliably identified 2423 counterparts out of the $\tilde Q_0
\times 6621 \approx 3925$ counterparts that are actually detected in
the SDSS DR7 data.

We show from a consideration of their sub-mm colours that those
sources without optical counterparts appear to reside at higher
redshifts than those with optical counterparts in our available
ancillary data. We compute the completeness of our reliable catalogue
as a function of redshift, and find that {\em Herschel}--ATLAS sources
with SDSS counterparts have a lower median redshift than the general
SDSS population, suggesting a bimodal $n(z)$ for {\em Herschel}--ATLAS
sources. For this bimodal $n(z)$, we find that the lower redshift
population has a median redshift of $z = 0.40^{+0.25}_{-0.19}$ (with
the errors calculated according to the 16th and 86th percentiles of
the redshift cumulative frequency distribution), and that the high
redshift population peaks at $z > 1$. We also find evidence for a
population of sub--millimetre--selected interacting galaxies, and
suggest a possible method for selecting samples of strongly--lensed
galaxies. Finally, we find five new positions for {\em
  IRAS}--FSC\slash IIFSC$z$ sources based on our LR analysis and
higher-resolution PACS and SPIRE data.

The UV\slash optical\slash near--infrared identifications to the
250\,$\mu$m--selected sample, as well as their photometric and
spectroscopic redshifts, are available for download from the {\em
  Herschel}--ATLAS webpage; \url{http://www.h-atlas.org}.

\section*{Acknowledgments}
{\em Herschel} is an ESA space observatory with science instruments
provided by European-led Principal Investigator consortia with
significant participation from NASA. U.S. participants in {\em
  Herschel}--ATLAS acknowledge support provided by NASA through a
contract issued from JPL. GAMA is a joint European-Australasian
project based around a spectroscopic campaign using the
Anglo-Australian Telescope. The GAMA input catalogue is based on data
taken from the Sloan Digital Sky Survey and the UKIRT Infrared Deep
Sky Survey. Complementary imaging of the GAMA regions is being
obtained by a number of independent survey programs including {\em
  GALEX} MIS, VST KIDS, VISTA VIKING, {\em WISE}, {\em
  Herschel}--ATLAS, GMRT and ASKAP providing UV to radio coverage. The
GAMA website is: \url{http://www.gama-survey.org/}. This work used
data from the UKIDSS DR5 and the SDSS DR7. The UKIDSS project is
defined in Lawrence et al. (2007) and uses the UKIRT Wide Field Camera
(WFCAM; Casali et al. 2007). Funding for the SDSS and SDSS-II has been
provided by the Alfred P. Sloan Foundation, the Participating
Institutions, The National Science Foundation, the U.S. Department of
Energy, the National Aeronautics and Space Administration, the
Japanese Monbukagakusho, the Max Planck Society and the Higher
Education Funding Council for England. The Italian group acknowledges
partial financial support from ASI contract I/009/10/0 `COFIS'.

\bsp

\label{lastpage}

\appendix
\section{Checking the identification process}
\label{sec:checks}

\subsection{{\em IRAS} sources}
\label{sec:iraschecks}

In the 9hr field SDP region, there are a total of 35 detections from
the Imperial {\em IRAS}--FSC Redshift Catalogue (IIFSC$z$, Wang \&
Rowan-Robinson, 2009, building on the {\em IRAS} Faint Source
Catalogue of Moshir et al. 1992), the majority with associated
optical\slash near--infrared positions of high reliability. By
matching our 250\,$\mu$m selected catalogue with the IIFSC$z$, and
comparing the results to our LR analyses, we can provide a first check
on the accuracy of our associations. There are 30 sources in the
IIFSC$z$ that have catalogue positions within 10.0 arcsec of the
250\,$\mu$m source positions. Each of the positions in the IIFSC$z$
catalogue for these sources matches an SDSS DR7 position within 2
arcsec, and has reliability $R > 0.80$.

\begin{figure*}
\centering
\subfigure{\includegraphics[width=0.31\textwidth]{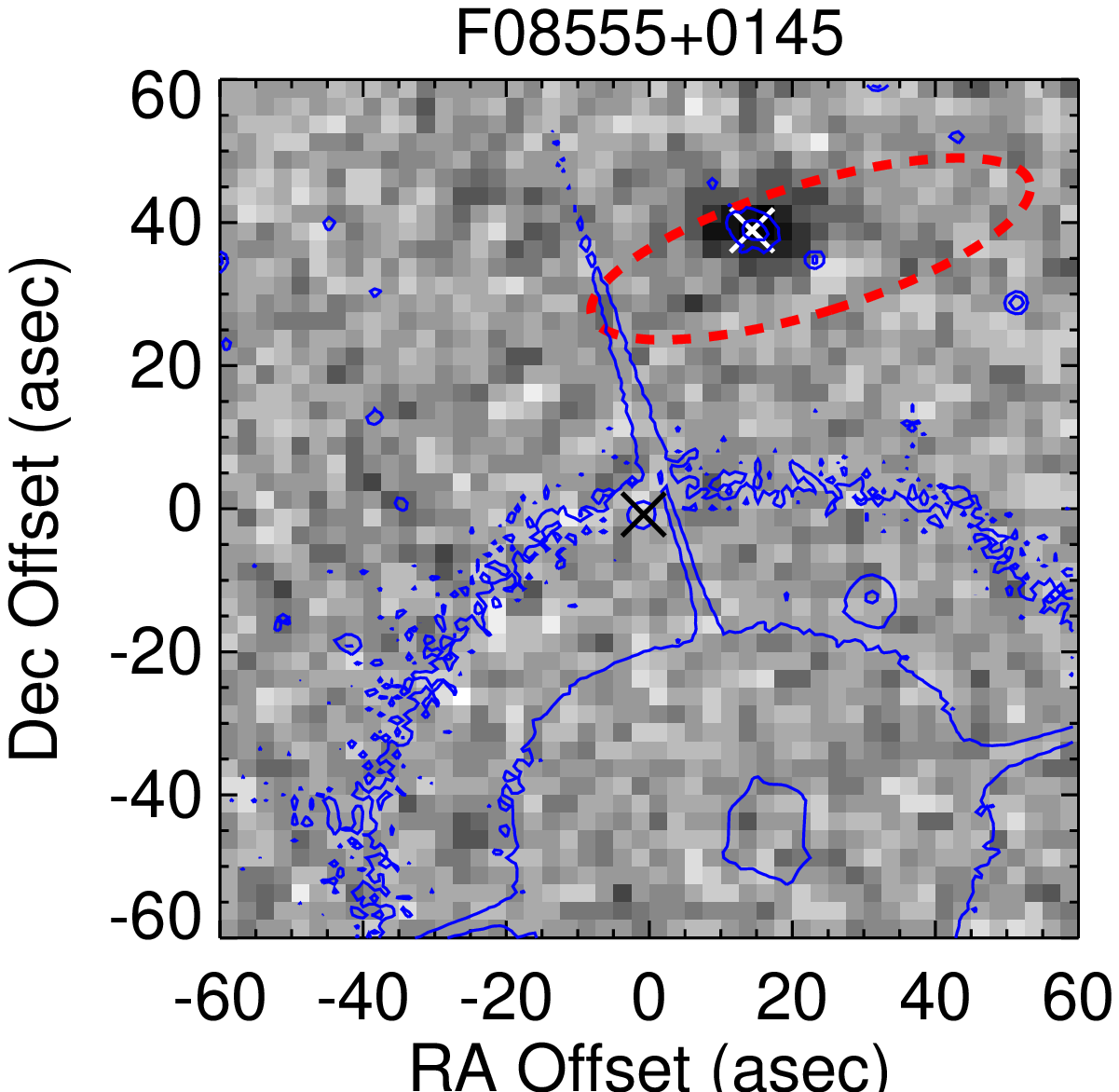}}
\subfigure{\includegraphics[width=0.31\textwidth]{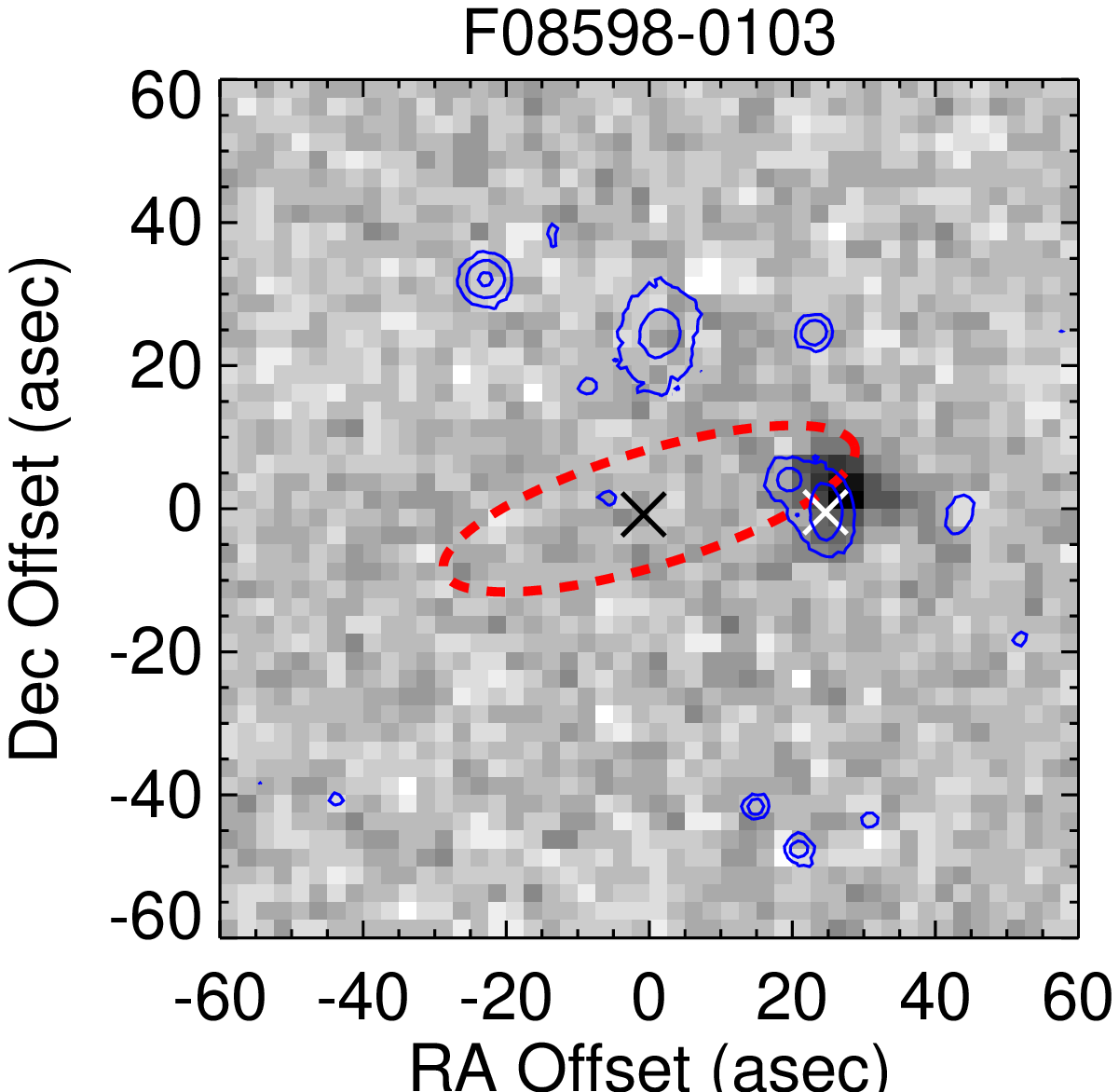}}
\subfigure{\includegraphics[width=0.31\textwidth]{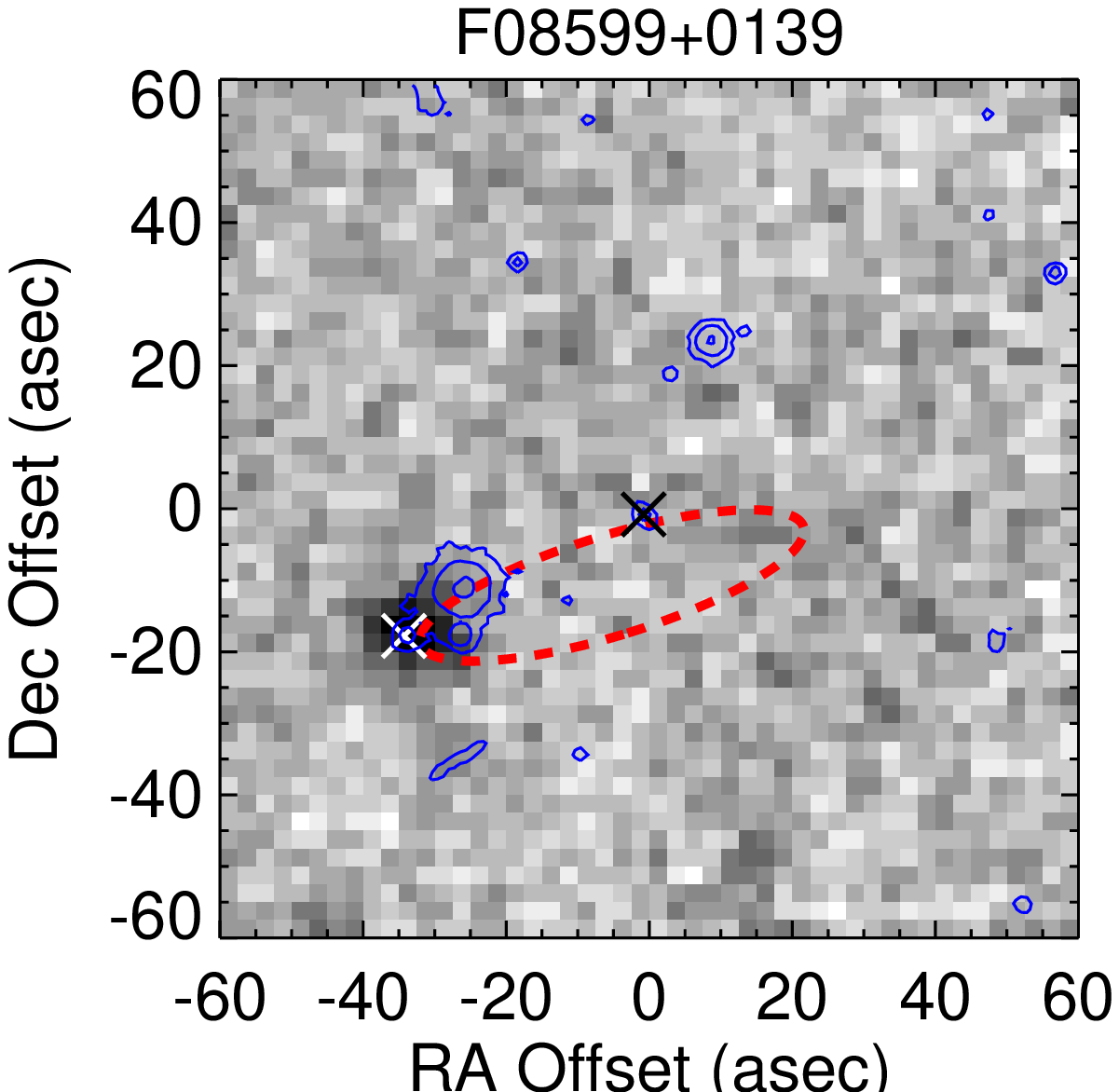}}
\subfigure{\includegraphics[width=0.31\textwidth]{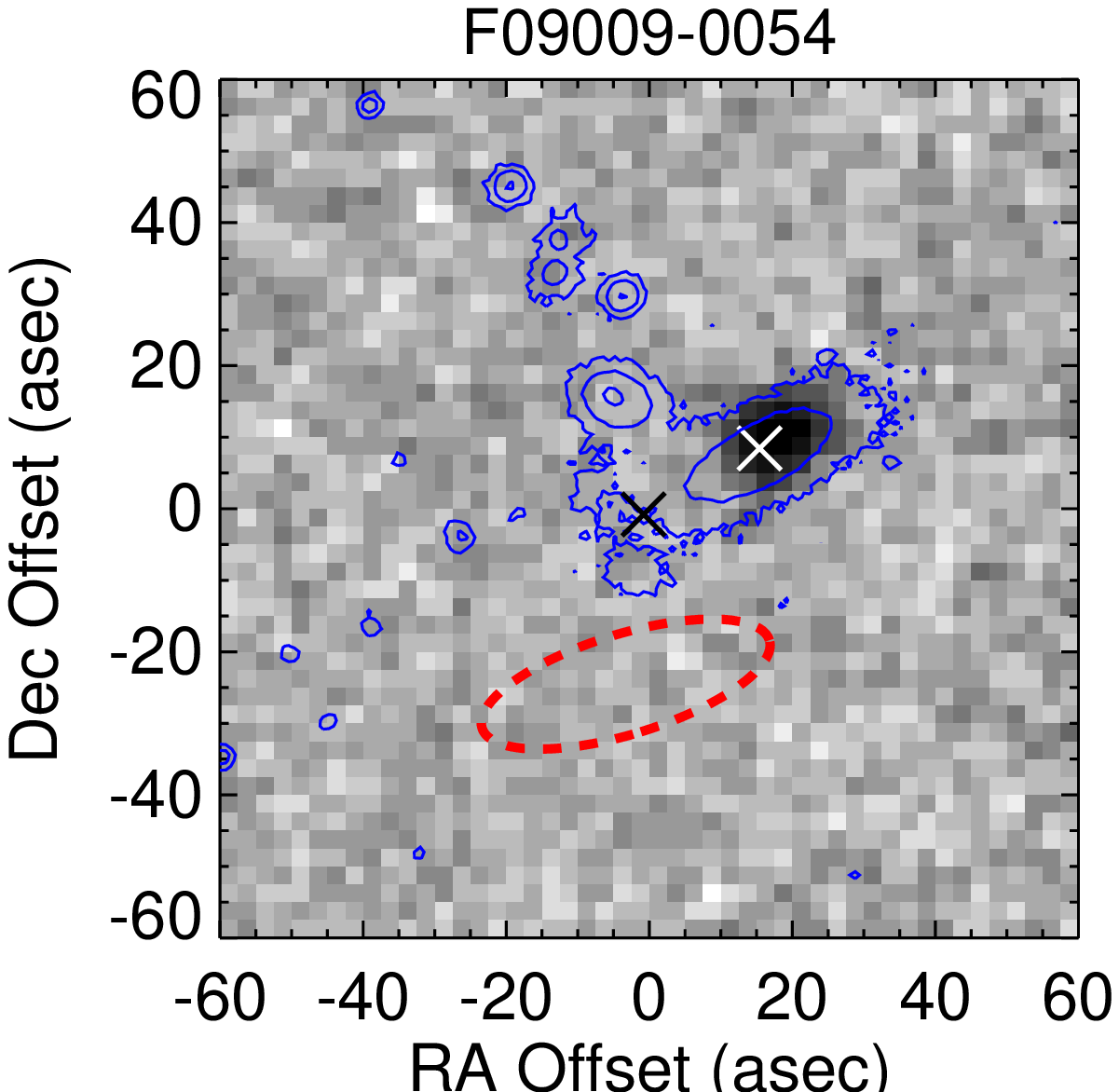}}
\subfigure{\includegraphics[width=0.31\textwidth]{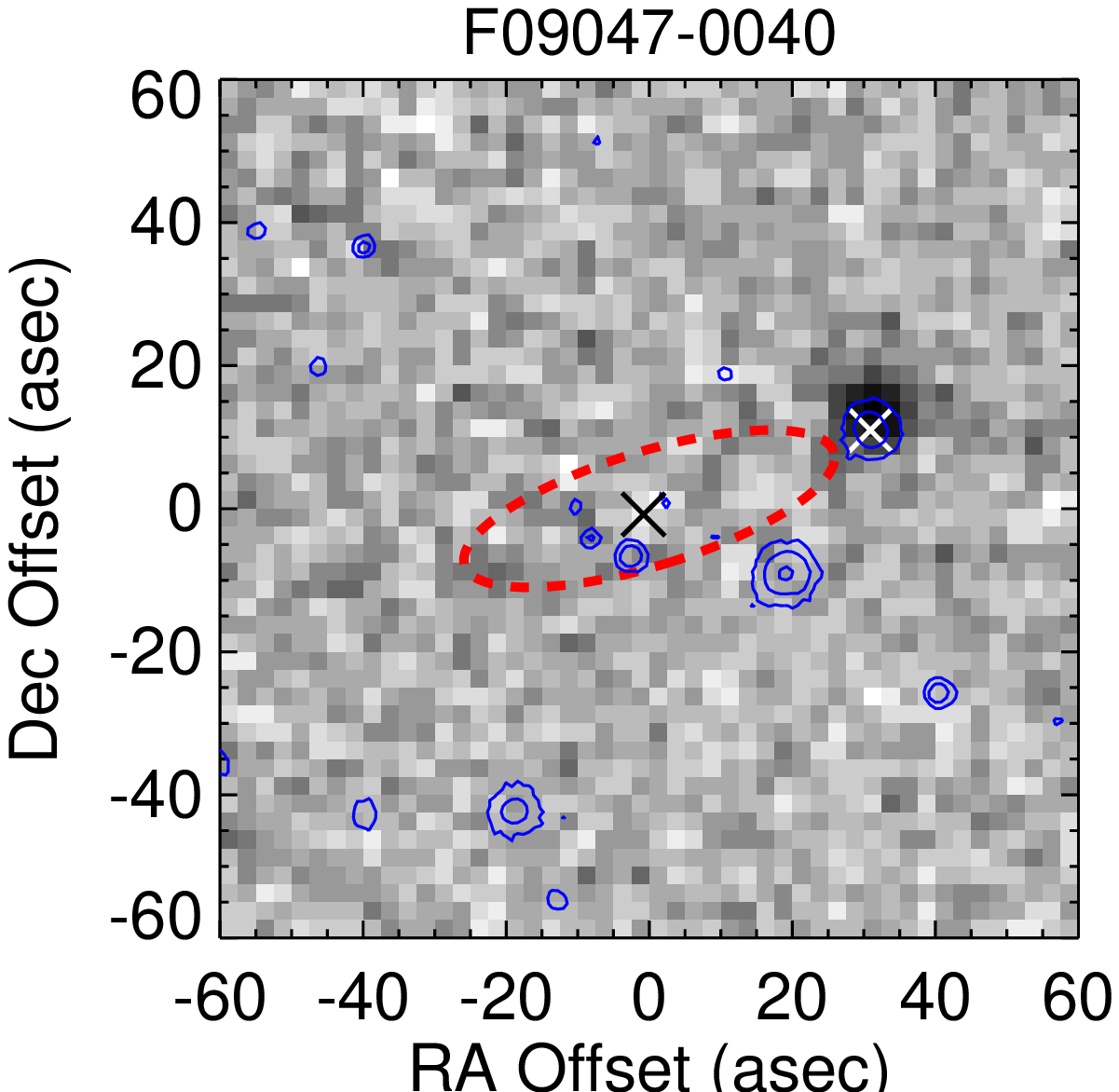}}
\caption{PACS 100\,$\mu$m greyscale cutout images showing the regions
  surrounding the five IIFSC$z$ catalogue positions for which we do
  not find a match within 10.0 arcsec in our SDSS\slash SPIRE
  catalogue. The IIFSC$z$ catalogue positions are denoted by a black
  cross (2 derived from SDSS positions, 1 from NVSS and 2 from the
  original FSC), with the SDSS DR7 $r$--band contours overlaid in blue
  and the {\em IRAS}--FSC 1$\sigma$ error ellipse overlaid in dashed
  red. The white crosses denote the positions of the $R \ge 0.8$ SDSS
  DR7 counterparts from our likelihood ratio analysis. These sources
  are discussed in more detail in section \ref{sec:iraschecks}. Using
  our SDSS DR7 likelihood ratio analysis and the higher-resolution
  SPIRE 250\,$\mu$m positions as our starting point, we are able to
  derive $R \ge 0.8$ counterparts for four of the five IIFSC$z$
  sources, positions of which are given in Table
  \ref{tab:iras_positions}. The exception is F08555+0145, for which
  the bright galaxy approximately centred on the PACS 100\,$\mu$m
  source is not present in the SDSS DR7 primary photometry catalogue
  (however we include a manually-measured position in Table
  \ref{tab:iras_positions}).  }
\label{fig:irasids}
\end{figure*}

There remain five IIFSC$z$ sources for which we do not recover
SDSS\slash SPIRE matches within 10.0 arcsec. In Figure
\ref{fig:irasids}, we show greyscale images of the {\em
  Herschel}--ATLAS PACS 100\,$\mu$m observations of the regions
surrounding these five IIFSC$z$ sources, centred on the quoted
IIFSC$z$ catalogue positions. It is clear that each IIFSC$z$ source
has a bright PACS detection less than one arcminute away, with the
PACS 100\,$\mu$m observations being of considerably higher sensitivity
and resolution than that of {\em IRAS} at 60\,$\mu$m (this is the band
on which the IIFSC$z$ is selected).

Here we discuss each of these sources individually.
\begin{itemize}
\item F08555+0145: This source has an IIFSC$z$ position derived from
  the SDSS DR6, residing approximately 40 arcsec away from the {\em
    IRAS} centroid. The bright PACS\slash SPIRE source within the
  1$\sigma$ positional errors of the IRAS centre is associated with an
  $r$ = 17.9\,mag galaxy approximately 1 arcsec away, which is not in
  the SDSS primary photometry catalogue. It is clear that this is the
  correct association, with a reliability based on its newly--measured
  magnitude and separation of $R = 0.999$, and that this source was
  mis-identified in the IIFSC$z$.
\item F08598-0103: IIFSC$z$ contains only an IRAS--derived position
  for this source in the absence of any counterparts detected in the
  ancillary data available at the time. Using our higher-resolution
  PACS\slash SPIRE observations, we are able to identify the optical
  counterpart, approximately 30 arcsec away from the IIFSC$z$
  position, with $R = 0.999$.
\item F08599+0139: The IIFSC$z$ position for this source is also
  derived from the SDSS DR6, suggesting a source approximately 10
  arcsec to the North of the IRAS position. {\em Herschel}--ATLAS
  data, however, reveal a bright sub-millimetre source $\sim$30 arcsec
  to the East, associated with an $r$--band counterpart at $R =
  0.994$. This source was therefore also mis-identified in the
  IIFSC$z$ catalogue.
\item F09009-0054: The IIFSC$z$ catalogue position for this source was
  derived using data from the NRAO VLA Sky Survey (NVSS, Condon et
  al., 1998), and resides within the extended stellar halo of the
  $z=0.04$ galaxy 2MASX J09033081-0106127 in the SDSS\slash
  UKIDSS--LAS data (which is also detected in each of the PACS and
  SPIRE bands). The higher resolution of the SDSS DR7 catalogue
  compared with the NVSS data and {\em IRAS} positions enables us to
  derive a more accurate position for the counterpart to this source,
  with $R = 0.999$.
\item F09047-0040: The PACS\slash SPIRE detection of this source is
  located approximately 40 arcsec away from the IRAS position quoted
  in the catalogue. We identify an SDSS DR7 optical counterpart with a
  more accurate position, and reliability $R = 0.999$.
\end{itemize}

Table \ref{tab:iras_positions} contains new positions for our reliable
counterparts to these {\em IRAS} sources. Assuming that our new
identifications to the {\em IRAS} sources are correct, we recover
reliable counterparts (with accurate positions) to all of the IIFSC$z$
sources at $R \ge 0.8$, as compared with $\sim$89 percent (31\slash
35) of sources for the IIFSC$z$ itself (we exclude the two sources
with clearly mis--identified counterparts, and also the two sources
with no identified counterparts).
 
\begin{table*}
\caption{Updated positions of the IIFSC$z$ sources, which were
  previously mis-identified, or identified with only NVSS\slash {\em
    IRAS} positions in Wang \&\ Rowan-Robinson (2009). The position
  angles of the {\em IRAS} positional error ellipses were orientated
  107$^\circ$ East of North.} \centering
%\scriptsize
\begin{tabular}{lllllllllllll}
\hline
 & \multicolumn{4}{l}{New positions} & \multicolumn{3}{l}{IIFSC$z$ catalogue positions} & \multicolumn{4}{l}{{\em IRAS}--FSC catalogue positions}\\
IIFSC$z$ ID & RA & Dec & source & $R$ & RA & Dec & source & RA & Dec & $\sigma_{\rm maj}$ & $\sigma_{\rm min}$ \\
\hline
F08555+0145 & 134.535 & 1.5649 & This paper & 0.999 & 134.539 & 1.5538 & SDSS DR6 & 134.533 & 1.5639 & 32\asec\ & 9\asec\ \\
F08598-0103 & 135.602 & $-$1.2622 & SDSS DR7 & 0.999 & 135.609 & $-$1.2623 & {\em IRAS}--FSC & 135.609 & $-$1.2623 & 30\asec\ & 8\asec\ \\ 
F08599+0139 & 135.636 & 1.4582 & SDSS DR7 & 0.994 & 135.627 & 1.4629 & SDSS DR6 & 135.628 & 1.4599 & 28\asec\ & 7\asec\  \\
F09009-0054 & 135.879 & $-$1.1033 & SDSS DR7 & 0.994 & 135.883 & $-$1.1059 & NVSS & 135.884 & $-$1.1127 & 21\asec\ & 7\asec\  \\
F09047-0040 & 136.829 & $-$0.8693 & SDSS DR7 & 0.999 & 136.838 & $-$0.8726 & {\em IRAS}--FSC & 136.838 & $-$0.8726 & 27\asec\ & 8\asec\ 
\label{tab:iras_positions}
\end{tabular}
\end{table*}

\subsection{A comparison with radio observations}
\label{radiocomp}

We also compared the results of our likelihood ratio analysis to data
from the Faint Images of the Radio Sky at Twenty centimetres (FIRST)
Survey (Becker, White \& Helfand, 1995). The FIRST survey covers 9,000
square degrees of sky with a resolution of 5 arcsec, with a source
density of approximately 90 per square degree brighter than the
detection threshold of 1\,$mJy$. At these relatively bright flux
limits, the source population is dominated by Active Galactic Nuclei
(AGN) rather than star--forming sources (e.g. Wilman et al., 2008); as
a result the overlapping population of sources between the {\em
  Herschel}--ATLAS and FIRST catalogues is not expected to dominate
the number counts.  

To make the comparison between our LR analysis and FIRST sources, we
used the frequentist identification procedure of Downes et al. (1986),
commonly used to quantify the formal significance of possible
counterparts to sub-millimetre galaxies in radio survey data
(e.g. Lilly et al., 1999, Ivison et al. 2007).  In this procedure, the
statistic used to assess the probability that a nearby radio source is
{\it not} associated with the SPIRE source is $S = \pi r^2 \times
n(>F)$, where $r$ is the angular distance between the SPIRE source and
the radio source, $F$ is the flux density of the radio source, and
$n(>F)$ is the surface density of radio sources with flux densities
greater than this. For each SPIRE source, we looked for radio sources
in the FIRST catalogue within 10.0 arcsec, and treated the radio
source with the lowest value of $S~(S_{\rm min})$ as the one most
likely to be associated with the SPIRE source. We used a Monte-Carlo
simulation (e.g. Eales et al. 2009) to determine the probability
distribution of $S_{\rm min}$ on the null hypothesis that there are no
genuine associations between radio sources and SPIRE sources.

We then used this probability distribution to determine the
probability that each measured value of $S_{\rm min}$ would have
occured by chance. We call this probability $P^\prime$. Of the 6621
$5\sigma$ 250\,$\mu$m SPIRE sources, 105 have radio counterparts
within 10.0 arcsec, all with values of $P^\prime < 0.002$. However,
this does not take account of the fact that with such a large sample
of SPIRE sources one expects to find some low values of $P^\prime$
even if there were no genuine associations between the SPIRE sources
and FIRST objects. We used a Monte-Carlo simulation to determine that
15 of the 105 associations are likely to be spurious. To correct for
this, we calculated a new probability for each association, $P =
\alpha P^\prime$, where $\alpha$ is a constant that we calculated
using $\sum_i \alpha P^\prime_i = 15$. We took the conservative
decision to treat associations with $P < 0.2$ as counterparts which
are likely to be genuine, which rejected 29 of the original 105
associations.

There were a total of 76 SPIRE sources with $P < 0.20$ counterparts,
and each of these was scrutinised using the FIRST and SDSS images
displayed side--by--side with the Downes et al. and LR analysis
overlaid. In this manner, we compared the results of the two
independant identification methods. In forty-two cases, the $P < 0.20$
radio counterpart is also identified as having $R \ge 0.80$ in the
$r$--band data, and the two methods choose the same counterpart.

There are thirty SPIRE sources with high quality ($P < 0.20$) FIRST
counterparts which we do not recover in our LR analysis, including
twenty--three SPIRE sources which do not have any $r$--band
counterparts in our SDSS DR7 data (presumably distant,
optically--faint radio sources). Of the remaining seven sources with
$P < 0.20$ FIRST counterparts:

\begin{itemize}
  \item Four counterparts are detected in the optical data but have
    low reliabilities due to their faint magnitudes, or large
    separations in comparison to the value of $\sigma_{\rm pos}$
    derived based on the 250\,$\mu$m source SNR.
  \item Two sources have multiple, possibly interacting components
    with $L > 10.0$ but $R < 0.8$, only one of which is a radio source
    (these sources are discussed in section \ref{mergers}).
  \item In one further instance, the radio source has a double-lobed
    structure (a so-called FR-II, following Fanaroff \& Riley, 1974),
    not coincident with either the dust emission or the starlight in
    the plane of the sky. The lobes of this FRII are extremely bright;
    as a result, the $P$ statistic suggests that there is a low
    probability of a chance association, even though the separation
    between the SPIRE position and the FIRST centroid is large. The LR
    technique identifies the apparent host galaxy -- aligned at the
    centre, between the two luminous radio jets -- as having $L = 0.0$
    due to its large separation ($\sim 9$ arcsec) from the SPIRE
    centroid; this is an example of the limitations of the Downes et
    al. method.
%  \item IIFSC$z$ source F08555+0145
\end{itemize}

\begin{figure*}
  \centering
  \subfigure[$P=0.08, R=0.87$]{\includegraphics[width=0.24\textwidth]{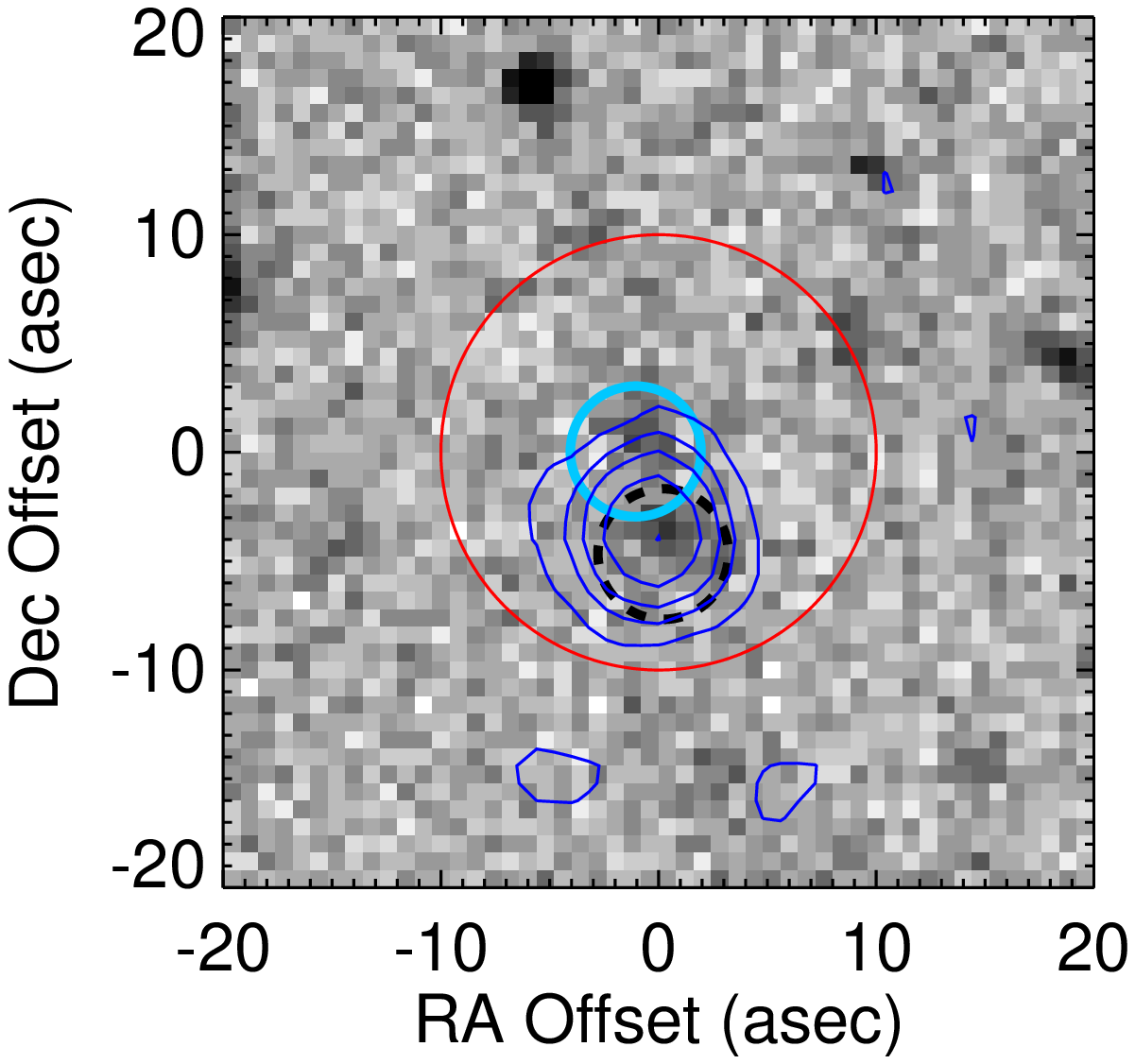}}
  \subfigure[$P=0.07, R=0.98$]{\includegraphics[width=0.24\textwidth]{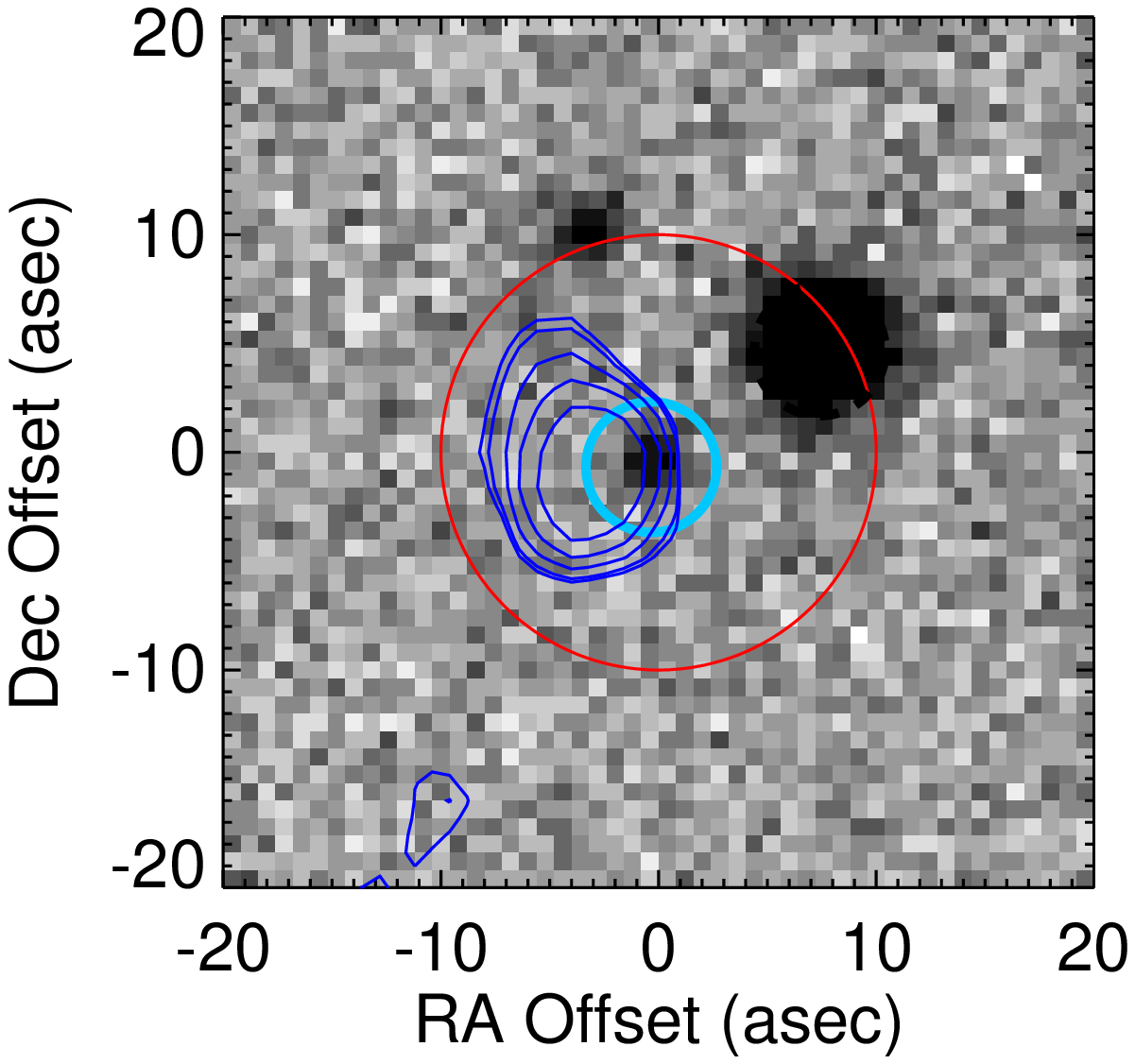}}
  \subfigure[$P=0.02, R=0.81$]{\includegraphics[width=0.24\textwidth]{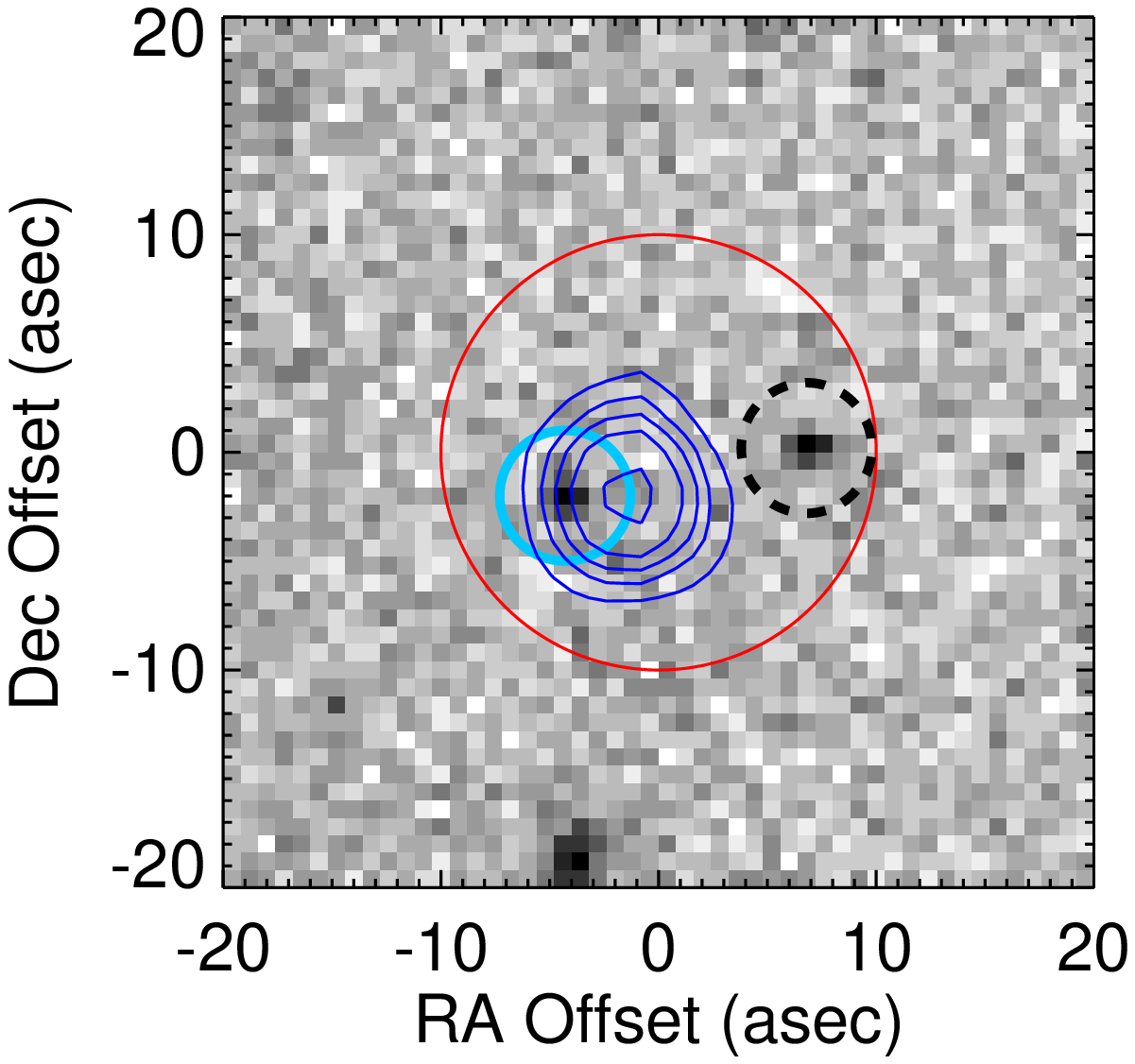}}
  \subfigure[$P=0.19, R=0.93$]{\includegraphics[width=0.24\textwidth]{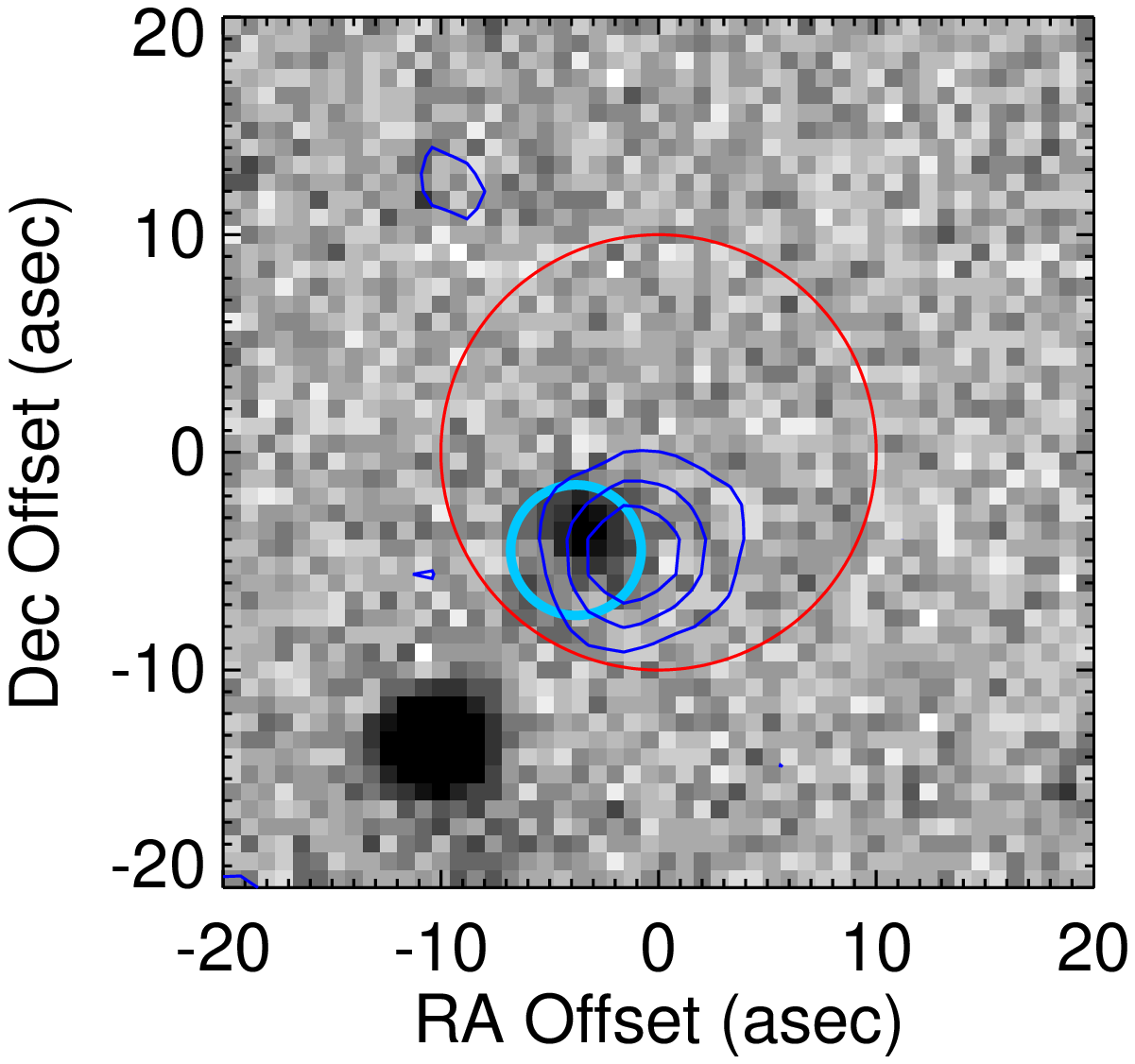}}
  \caption{Cases in which the LR method applied to the SDSS DR7 data
    and the P--statistic (Downes et al. 1986) method applied to the
    FIRST data produce different robust counterparts to 250\,$\mu$m
    sources. In each panel, the 10.0 arcsec search radius around the
    250\,$\mu$m position is shown in red, with any unreliable
    counterparts circled in black. Reliable counterparts from the LR
    analysis are circled in light blue, while the royal blue contours
    reveal the FIRST counterpart. The $P$ statistic for the FIRST
    source and the value of the reliability, $R$, of the most reliable
    SDSS DR7 counterpart are given in the subfigure captions for each
    250\,$\mu$m object. Thumbnail images are orientated such that
    North is up and East is to the left.}
  \label{fig:downes_lr_dis}
\end{figure*}

However, these possibilities do not contaminate the 250\,$\mu$m
selected sample with incorrect associations. There are however, four
instances where distinct counterparts have $P > 0.20$ and $R \ge
0.80$; here, the opposite is potentially true and the two methods
conflict. These sources have derived reliabilities of 0.87, 0.98, 0.81
and 0.93 as compared with distinct Downes et al. counterparts with $P$
statistics of 0.08, 0.07, 0.02 and 0.19, respectively. These sources
are shown in Figure \ref{fig:downes_lr_dis}, in which the 10.0 arcsec
search radius centred on the 250\,$\mu$m position is shown in red, any
unreliable optical counterparts in black, the reliable optical ID in
light blue, and the radio contours overlaid in royal blue. In two of
the four cases, the additional sources implied by the radio data are
visible in $K_S$--band observations from VIKING (Sutherland, 2009),
indicating that these sources are not merely effects of the larger
positional uncertainty in FIRST as compared with SDSS. Futhermore,
three of the four sources have SPIRE colours $S_{250} \slash S_{350}
\le 1.5$, suggesting high redshifts ($z > 1$) or cold dust
temperatures, with the former being at odds with the photometric
redshifts of their most reliable counterparts ($z < 0.55$). Sources
with similar SPIRE colours and low-redshift counterparts are discussed
in more detail in section \ref{submmcolours}.

%2978: zphot = 0.54+/-0.08
%4213: zphot = 0.44+/-0.12
%6345: zphot = 0.50+/-0.07
%6714: zphot = 0.41 +/- 0.03

Finally, we note that probabilistic arguments such as those discussed
here will inevitably present apparent disagreements for a small number
of sources within large samples. In the remaining 101 out of 105 cases
however, the results of our LR analysis are consistent with those
using the FIRST catalogue and the $P$ statistic, and crucially we
recover an additonal 2,348 counterparts, compared with 31 extra
counterparts to the 250\,$\mu$m sources obtained by using only the
radio data.

\subsection{{\it Spitzer} observations}
\label{spitzer}

An additional check on the identification process was conducted by
searching for mid--infrared data from the {\it Spitzer Space
  Telescope} heritage archive, in order to compare the reliabilities
from our $r$--band catalogue with near- and mid--infrared images
between 3.6 and 160\,$\mu$m. Four sets of observations were found
which overlapped with the {\em Herschel}--ATLAS SDP observations
%(programme IDs 1307, 1724, 1861,
%20410, 30344, 40640, 50740, with PIs Reach, Noriega-Crespo,
%Noriega-Crespo, van Gorkom, Jarvis, Croft, \&\ Fazio,
%respectively). 
These data can be used to examine the regions surrounding the SPIRE
IDs for additional sources which may not be present in the $r$--band
catalogue used for the identification process, as a visual check on
the effectiveness of the LR technique.

\begin{figure*}
\centering
\includegraphics[width=0.99\textwidth]{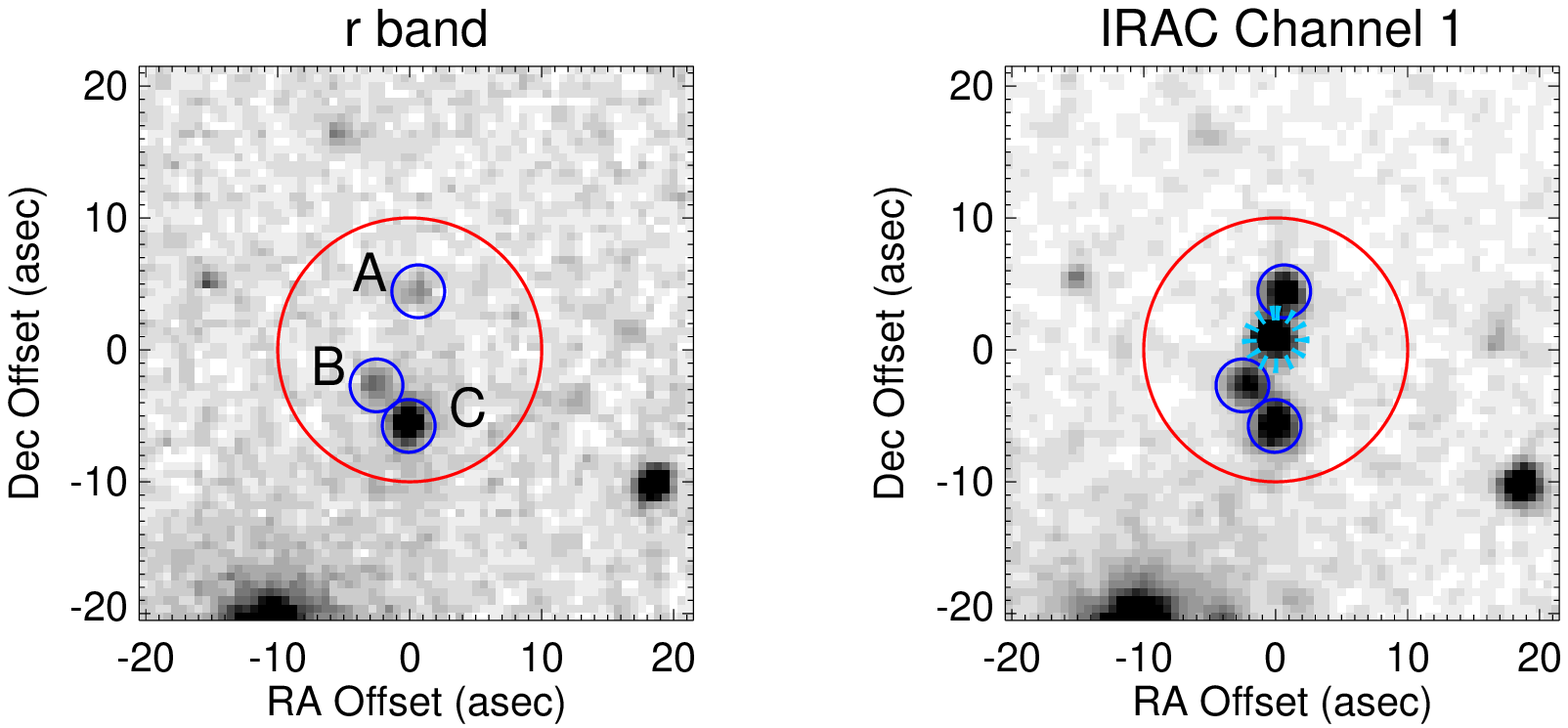}
\caption{SDSS $r$--band (left) and {\it Spitzer Space Telescope} IRAC
  3.6\,$\mu$m image of the region surrounding 
  %SPIRE ID 65 
  source H--ATLAS J090913.2+012111, at $\alpha = $137.305, $\delta =$
  1.3532 (position shown by the red circle, which has a radius of 10.0
  arcsec). The $r$--band image contains three sources (blue 2 arcsec
  circles) that are identified as potential counterparts to the SPIRE
  source, with reliabilities of 0.00, 0.29, and 0.00 (and $L = 0.00,
  0.17$ and 0.00) for the sources labelled A, B and C,
  respectively. The IRAC 3.6\,$\mu$m channel image (right) reveals an
  additional source within 1 arcsec of the SPIRE centroid (dotted
  light blue 2 arcsec radius circle). The low reliabilities associated
  with the $r$--band detections indicates the power of the LR
  technique in this context. Images are orientated such that North is
  up and East is to the left.}
\label{fig:rband_irac1}
\end{figure*}

There are a total of 49 sources that have {\it Spitzer} data, and
although these data vary in sensitivity, there is no evidence that
would suggest a mis-identification from the $r$--band catalogue. Such
indications of wrong IDs would include reliable ($R \ge 0.8$)
$r$--band counterparts indicated for SPIRE sources which have
previously unrevealed bright {\em Spitzer} sources nearer to the
centre of the SPIRE centroid. Indeed, in one case in particular
(H--ATLAS J090913.2+012111), the sensitive IRAC data reveal the power
of the LR technique. Although there are three potential counterparts
in the SDSS DR7 $r$--band catalogue all within 6 arcsec of the SPIRE
centroid, they have all been given low reliability ($R \le 0.30$, and
also $L \le 0.20$). The IRAC 3.6\,$\mu$m data reveal a fourth
candidate counterpart within 1 arcsec of the SPIRE position, which is
presumably the true counterpart. The $r$--band and IRAC 3.6\,$\mu$m
data are presented in Figure \ref{fig:rband_irac1}, with the various
source positions overlaid to demonstrate the robustness of the LR
method for this particular source, but also the need for
longer-wavelength observations in order to be able to reliably
identify the counterparts to higher redshift sources. The forthcoming
data from the VISTA Kilo-degree INfrared Galaxy (VIKING) survey and
from the Wide-field Infrared Survey Explorer ({\it WISE} -- Duval et
al., 2004) satellite will enable this.

This example also highlights one crucial advantage of using the LR
technique for {\em Herschel} surveys rather than opting simply for the
Downes method; the LR method takes into account the fact that not
every source has a counterpart that is brighter than the detection
limit in ancillary survey data.

\end{document}